\documentclass[a4paper,11pt]{article}
\usepackage{jcappub} 
\usepackage{lineno}
\usepackage{subfig}

\usepackage{booktabs}
\usepackage{multirow}
\usepackage{soul}

\title{\boldmath  A gamma-ray study of Galactic PeVatron candidates LHAASO J1825-1326 and LHAASO J1839-0545}

\author[a,1]{Rubens Jr. Costa, \note{Corresponding author.}}
\author[a,c]{Débora B. Götz,}
\author[a,b,c]{Rita C. Anjos,}
\author[d]{Luiz. A. Stuani Pereira,}
\author[a]{Alexandre J. T. S. Mello}

\affiliation[a]{Programa de Pós-Graduação em Física e Astronomia, Universidade Tecnológica Federal do Paraná (UTFPR), Av. Sete de Setembro, 3165, 80230-901, Curitiba, PR, Brazil}
\affiliation[b]{Departamento de Engenharias e Exatas, Universidade Federal do Paran\'a (UFPR), Pioneiro, 2153, 85950-000 Palotina, PR, Brazil}
\affiliation[c]{Programa de p\'os-graduação em F\'isica \& Departamento de F\'isica, Universidade Estadual de Londrina (UEL), Rodovia Celso Garcia Cid Km 380, 86057-970 Londrina, PR, Brazil}
\affiliation[d]{Programa de Pós-Graduação em Física Aplicada, Universidade Federal da
Integração Latino-Americana, 85867-670, Foz do Igua\c{c}u, PR, Brazil}
\affiliation[e]{Unidade Acadêmica de Física, Universidade Federal de Campina Grande (UFCG), R. Aprígio Veloso, 882, 58429-900, Campina Grande, PB, Brazil}

\emailAdd{rubensp@utfpr.edu.br; deboragotz@alunos.utfpr.edu.br; ritacassia@ufpr.br; luizstuani@uaf.ufcg.edu.br; ajmello@utfpr.edu.br}

\abstract{Recent studies by LHAASO have shown the presence of high-luminosity PeVatrons in our Galaxy. We examine two notable sources, each consisting of two pulsars, detected by LHAASO. We study gamma and
cosmic-ray emissions from the surroundings of these pulsars. 
We used the Gammapy software to perform gamma-ray measurements from these sources in anticipation of future analyses to be carried out with the CTAO, which is now under development. Furthermore, the particle propagation throughout the Galaxy was simulated using the GALPROP software, accounting for emission from energy losses due to spin-down. As a result, we present the particle spectra generated during this propagation phase along with the corresponding gamma-ray emission.  
 The findings suggest that CTAO may have the capability to detect these sources and show that the data can be explained by gamma rays coming from a leptonic model.

}

\begin{document}
\maketitle
\flushbottom
\section{Introduction} \label{sec:intro}

Cosmic rays (CR) are accelerated to high energies and travel at nearly the speed of light, striking the Earth's atmosphere continuously. However, there is a lack of clear understanding about the origins of these particles, including how they undergo acceleration and their importance beyond our Galaxy \cite{ANCHORDOQUI20191}. In recent decades, cosmic-ray astrophysics entered an era of precision measurements. New observatories and analysis methodologies generated data of unprecedented quality. Nevertheless, the precise underlying mechanisms responsible for the knee formation in the cosmic-ray spectrum and the boundary separating galactic and extragalactic sources remain elusive. The phenomenon known as the ``knee" is a prominent characteristic observed in the all-particle spectra of CR. It is characterized by a change in the spectral index around $10^{15}$ eV \cite{ANTONI20051}. At lower particle energies, specifically those below the knee, protons are the primary constituents of CR. However, as CR energy increases, the presence of heavier components becomes crucial. Based on the prevailing consensus, it is widely accepted that the phenomenon of spectral steepening found above the knee can be attributed to the diminished efficacy of Galactic cosmic-ray sources in facilitating the acceleration of particles to such high energies \cite{doi:10.1142/S0218271819300222,AbdulHalim_2023}.

Interactions between CR and cosmic environment particles (hadronic interactions) produce charged and neutral pions. The neutron pion decays into two photons, while the decay of charged pions produces  $e^{+}$ $\&$ $e^{-}$ – and neutrinos. The three fundamental mechanisms that can produce gamma-ray fluxes associated with CR are: a) the decay of neutron pions; b) electron/positron bremsstrahlung, and c) inverse Compton scattering of a photon by an electron. The aforementioned cascade processes demonstrate a positive correlation between the maximum value of the integral GeV - TeV gamma-ray flux and the maximum value of the luminosity of CR \cite{Supanitsky_2013, Anjos_2014, Sasse_2021, 2021JCAP...10..023D, Coelho_2022}. This correlation serves as a driving force for conducting gamma rays investigations to acquire insights into the origins of CR. 

In the present scenario, the Galactic PeVatrons\footnote{The term ``PeVatron" is used to describe a Galactic cosmic ray source that possesses the capability to accelerate particles to energies in the PeV range (equivalent to $10^{15}$ eV) and, correspondingly, generates gamma-rays exceeding 100 TeV.}, as seen by
ground-based gamma-ray facilities such as
the Large High Altitude Air Shower Observatory (LHAASO), gamma-ray emission at 100 TeV or higher has also been observed by the High-Altitude Water Cherenkov (HAWC) observatory, and others have demonstrated their capacity to serve as a fertile setting for elucidating the characteristics of particles that pertain to the knee area and exhibit emissions in the high energy spectrum. LHAASO is located in the province of Sichuan, China, at an elevation of 4410 meters. It comprises a collection of extended air shower (EAS) detectors developed explicitly to investigate CR and gamma rays within the energy range of sub-PeV to 1000 PeV \cite{Zhen2010}. The system consists of three components: the Wide Field-of-view Cherenkov Telescope Array (WFCTA), the Water Cherenkov Detector Array (WCDA), and the Kilometer Square Array (KM2A). The continuous construction of the LHAASO-KM2A has enabled the identification of 43 Galactic PeVatrons categorized as Ultra-High-Energy (E $>$ 100 TeV) \cite{Cao_2024}. Unsurprisingly, one can locate potential counterparts for the gamma-ray production regions and the nearby particle accelerators close to the extended ultra-high energy (UHE) sources. The list includes candidates such as pulsars and pulsar wind nebulae, supernova remnants, and young massive star clusters that may be responsible for the electron and proton PeVatrons in the Milky Way. The limited angular resolution of the LHAASO devices indicates that, on average, there are at least two potential TeV counterparts for each PeVatron \cite{Cao2021a}. The Crab Nebula, one of the largest non-thermal sources in our Galaxy and a representative of pulsar wind nebulae, is the only source firmly identified. Furthermore, the LHAASO has successfully identified the most energetic photon ever recorded, measuring at an unprecedented energy of 1.4 PeV. The spectral energy distributions (SEDs) of three bright sources, namely LHAASO J1825-1326, LHAASO J1908+0621, and LHAASO J2226+6057, were also published by LHAASO. 

In the future, one of the key scientific goals of the Cherenkov Telescope Array Observatory (CTAO) is to unveil the source of CR particles and neutrinos \cite{KSP}. This objective is related to the search for enigmatic powerful PeVatrons and ultra-high-energy cosmic-ray (UHECR) outside the Galaxy. CTAO will be a state-of-the-art gamma-ray observatory with almost 70 imaging atmospheric Cherenkov telescopes (IACTs). CTAO will be able to conduct comprehensive sky observations, utilizing both multi-wavelength and gamma-ray approaches. The potential utilization of CTAO for observing PeVatrons sources holds promise in elucidating the correlation between CR and TeV gamma-rays \cite{KSP}. 

In this work, we focus on the PeVatrons LHAASO J1825-1326 and LHAASO J1839-0545 \cite{Cao2021a}. The first has the pulsars PSR J1826-1334 and PSR J1826-1256 as possible counterparts, and the last one has the pulsars PSR J1837-0604 and J1838-0537. We use the gamma-ray emissions to investigate the presence of cosmic rays in these pulsars' environment. We derived the gamma-ray spectrum from the source regions  (LHAASO J1825-1326 and LHAASO J1839-0545) and evaluated CTAO performance using Gammapy\footnote{https://gammapy.org/} \citep{Deil2017} software. The Gammapy software is an open-source Python package for gamma-ray analysis built on Numpy\footnote{https://numpy.org/}, Scipy\footnote{https://scipy.org/}, and Astropy\footnote{https://www.astropy.org/}.
It is a core library for the Science Analysis Tool of CTAO. It provides tools to simulate the gamma-ray sky for a variety of telescopes, such as CTAO, High Energy Stereoscopic System (HESS) \cite{HESS}, Very Energetic Radiation Imaging Telescope Array System (VERITAS) \cite{Veritas}, Major Atmospheric Gamma Imaging Cherenkov (MAGIC) \cite{MAGIC}, HAWC \cite{2014AdSpR..53.1492W}, and Fermi large area telescope (Fermi-LAT) \cite{Atwood2009}. Additionally, the GALPROP\footnote{https://galprop.stanford.edu/} software \cite{Porter_2022} was employed to examine the signals of the propagation of CR particles within the Galaxy. GALPROP is a numerical code that simulates galaxy parameters and source properties to solve the transport equation using the second-order Crank Nicolson method. The program calculates the propagation and cosmic ray interactions encompassing gamma-ray phenomena such as Bremsstrahlung, Inverse Compton scattering, and pion decay. Energy losses and fragmentation are based on realistic distributions of the radiation fields and interstellar gas \cite{Johannesson_2018}. The GALPROP program is described in depth in \cite{Strong_1998,Moskalenko_1998,Strong_2000,Porter_2022}.
 
The results demonstrate the effectiveness of CTAO in observing these sources in the future, as well as the strong agreement of gamma-rays from CR in describing the high-energy data. The structure of this paper is as follows. The details of the sources examined are presented in Section \ref{sec:2}. We report the Gammapy results in Section \ref{sec:3}. The GALPROP gamma-ray spectra are described in Section \ref{sec:4}; finally, our findings are summarized in Section \ref{sec:5}.

\section{Galactic PeVatron candidates LHAASO J1825-1326 and LHAASO J1839-0545} \label{sec:2}

The potential PeVatron LHAASO J1825-1326 is one of the three most significant ($\rm{16.4\ \sigma}$ at $ \rm{E > 100\ TeV}$) Galactic gamma-ray sources (the other two being the Crab and LHAASO J1908+0621) as published by the LHAASO Collaboration \cite{Cao2021a}. The SED exhibits a pronounced peak at $\rm{100\ TeV}$. It gradually steepens with energy between $\rm{10\ TeV}$ and $\rm{500\ TeV}$, possibly due to photon-photon absorption that occurs during their interactions with the diffuse far-infrared and microwave radiation fields. The LHAASO has demonstrated that spectral fits can be better described by the logarithmic parameter function given as $E_{\rm{10TeV}}^{(-0.92 -1.19 \log E_{\rm{10TeV}})}$, where $E_{\rm{10TeV}} \equiv E/ 10\ \rm{TeV}$,  rather than a power-law model, ${E^{-3.36}}$, with the Akaike Information Criterion \cite{Akaike1974} ($\rm{AIC}$) giving $\rm{AIC_{LOG} = 11.6}$ and $\rm{AIC_{PL}} = 14.8$. 
LHAASO J1825-1326 is proposed to have originated from either PSR J1826-1334 or PSR J1826-1256. Potential TeV counterparts are suggested to include HESS J1826-130, 2HWC J1825-134, and HESS J1825-137. The characteristics and separation from the LHAASO source of the pulsars PSR J1826-1334 and PSR J1826-1256 are given in Table \ref{tab:pulsars_char}.
\begingroup
\renewcommand*{\thefootnote}{\alph{footnote}}  
\begin{table}[htbp]
\centering
\begin{minipage}{\textwidth}
    \centering
    \begin{tabular}{lll}
    \toprule
         Pulsar & PSR J1826-1334 & PSR J1826-1256\\
         \hline     
\midrule
         spin-down age $\rm \tau\ (yr)$ & $\rm{2.14 \times 10^{4}}$ & $\rm{1.44 \times 10^{4}}$\\
         surface dipole magnetic\\
         field strength $\mathrm{B_{s}}\ (\mathrm{G})$ & $\rm{2.8 \times 10^{12}}$ & $\rm{3.7 \times 10^{12}}$\\
         spin-down power $\rm{\dot E\ (erg\ s^{-1}})$ & $\rm{2.8 \times 10^{36}}$ & $\rm{3.6 \times 10^{36}}$\\
         period $\rm P \ (s)$ & $\rm{0.1015}$ & $\rm{0.1102}$\\
         location R.A., Dec. (deg.) & $\rm{276.55, -13.58}$ & $\rm{276.54,-12.94}$\\
         distance (kpc) & $\rm{\approx 3.6}$ & $\rm{\approx 1.55}$\\
         separation from \\
         LHAASO J1825-1326 (deg.) & $\rm{\approx 0.2}$ & $\rm{\approx 0.5}$\\
    \bottomrule
    \end{tabular}
    \caption[]{Characteristics of PSR J1826-1334 \cite{Manchester2005}, also referred to as PSR B1823-13, \cite{Clifton1986,Yuan2010, Yao_2017} and PSR J1826-1256 \cite{Abdo2009, Ray2011} in the region of the LHAASO J1825-1326, with the $68\%$ containment radius  of $0.62^{\circ}$ \cite{Cao2021a}.}    
    \label{tab:pulsars_char}
    \end{minipage}
\end{table}
\endgroup

LHAASO J1839-0545 stands out as another potential PeVatron, having been identified by the LHAASO \cite{Cao2021a} with a significance of $\rm{7.7\ \sigma}$ at energies greater than $\rm{100\ TeV}$. This source boasts a maximum energy of $\rm{E_{max} \approx 210\ TeV}$ and exhibits a differential photon flux of $\rm{0.70 \pm 0.18\ CU}$ at $\rm{100\ TeV}$ \footnote{CU, flux of the Crab Nebula at $\rm{100\ TeV;\ 1\ CU = 6.1 \times 10^{-17}\ photons\ cm^{-2}\ s^{-1}\ TeV^{-1}}$.}. Possible candidate pulsars associated with this source include PSR J1837-0604 and PSR J1838-0537. HESS J1837-069, 2HWC J1837-065, and HESS J1841-055 have been suggested as potential TeV counterparts. Details about the pulsars PSR J1837-0604 and PSR J1838-0537 can be found in the Tab \ref{tab:pulsars_char2}.

\begingroup
\renewcommand*{\thefootnote}{\alph{footnote}}  
\begin{table}[htbp]
\centering
\begin{minipage}{\textwidth}
    \centering
    \begin{tabular}{lll}
    \toprule
         Pulsar & PSR J1837-0604 & PSR J1838-0537\\
         \hline     
\midrule
         spin-down age $\rm \tau\ (yr)$ & $\rm{3.38 \times 10^{4}}$ & $\rm{4.89 \times 10^{3}}$\\
         surface dipole magnetic\\
         field strength $\mathrm{B_{s}}\ (\mathrm{G})$ & $\rm{2.1 \times 10^{12}}$ & $\rm{8.4 \times 10^{12}}$\\
         spin-down power $\rm{\dot E\ (erg\ s^{-1}})$ & $\rm{2.0 \times 10^{36}}$ & $\rm{6.0 \times 10^{36}}$\\
         period $\rm P\ (s)$ & $\rm{0.0963}$ & $\rm{0.1457}$\\
         location R.A., Dec. (deg.) & $\rm{279.43, -6.08}$ & $\rm{279.73, -5.62}$\\
         distance (kpc) & $\rm{\approx 4.8}$ & $\rm{\approx 1.3}$\\
         separation from\\
         LHAASO J1839-0545  (deg.) & $\rm{\approx 0.6}$ & $\rm{\approx 0.3}$\\
    \bottomrule
    \end{tabular}
    \caption[]{Characteristics of PSR J1837-0604 \cite{Manchester2005, DAmico_2001,Morris2002} and  PSR J1838-0537 \cite{Pletsch_2012} in the region of the LHAASO J1839-0545, with the angular extension $\leq 1^{\circ}$ \cite{Cao2021a}$^{a}$.
    }    
    \footnotesize{$^{a}$ For the angular extension of this source, the authors only provided an upper bound.}
    \label{tab:pulsars_char2}
    \end{minipage}
\end{table}
\endgroup

\section{Sensitivity of CTAO to gamma-ray emission}
\label{sec:3}

We lack precise information regarding the origin of the LHAASO observations, but it is known that two TeV counterparts (pulsar-associated nebulae) surround each of them.  In this section, we aimed to investigate whether CTAO observations could aid in identifying the sources of the LHAASO observations. This may be achieved by simulating CTAO observations in the multi-GeV to multi-TeV range within the two counterpart regions while considering the LHAASO observation in both scenarios. Using this approach,  we estimate the source's spectrum by combining data from the Fermi-LAT, HESS, HAWC, and LHAASO instruments and performing a simultaneous likelihood fit. An exponential cutoff power law (ECPL) was considered for the spectral shape, and it is defined as $\phi(E) = \phi_0 \cdot {E_{\rm{10TeV}}}^{ -\Gamma} \exp(-{E}/{E_{\rm{cut}}})$, where $\rm{\phi_0}$ is the amplitude, $\rm{\Gamma}$ is the index, and ${E_{\rm cut}}$ is the cutoff energy. We only use the spectral model in the simulations because it is the mandatory component. 
 
To evaluate the capability of CTAO to detect gamma-ray emissions coming from these regions, we employed the 1D ON/OFF observation technique  \citep{Piano(2021)}. The ON/OFF method, a well-established technique in Cherenkov astronomy for 1D spectral extraction, involves using the aperture photometry tool to measure the source's emission \citep{Piano(2021)}. In this method, the background is determined from the data by observing an OFF region (where no sources are present), and then the photon count ($\rm{N_{OFF}}$) is extracted from that region. The photon count ($\rm{N_{ON}}$) is obtained by the ON region, centered around the source. The ON and OFF regions are characterized by different effective areas (A), exposure times (t), and region sizes (k). The photon excess is calculated as $\rm{N}_{\rm{S}} = \rm{N}_{\rm{ON}} - \alpha \rm{N}_{\rm{OFF}}$, where $\alpha$ represents the background scaling factor defined by:

\begin{equation}
\alpha = \frac{\rm{A}_{\rm{ON}} \cdot \rm{t}_{\rm{ON}} \cdot \rm{k}_{\rm{ON}}}{\rm{A}_{\rm{OFF}} \cdot \rm{t}_{\rm{OFF}} \cdot \rm{k}_{\rm{OFF}}}.
\label{alpha}
\end{equation} 
The detection significance is computed via the analytic Li\&Ma formula \citep{LiMa(1983)}:

\begin{equation}
\rm{S} = \sqrt{2} \left\lbrace \rm{N}_{\rm{ON}} \rm{ln} \left[ \frac{1+\alpha}{\alpha}\left( \frac{N_{\rm{ON}}}{N_{\rm{ON}}+N_{\rm{OFF}}} \right) \right] + N_{\rm{OFF}} \rm{ln} \left[ (1+\alpha)\left( \frac{N_{\rm{OFF}}}{N_{\rm{ON}}+N_{\rm{OFF}}}\right) \right] \right\rbrace ^{1/2}. 
\label{LiMa}
\end{equation} 
In this work, the $\rm{N_{OFF}}$ value was estimated through the background template stored in the instrument response functions (IRFs) from the CTA Consortium and Observatory (version prod5 v0.1 \cite{CTAIRFS2021}), and the value of  $\alpha$  used is  0.2 (see the eq.~\ref{alpha}), based on the procedures for assessing CTAO sensitivity. The ON region was chosen by taking a simple circular region with the center and radius found to match those found in the HESS data  \cite{Abdalla2018, Abdalla2019} to the position and size of the pulsar-associated nebula inside the region. The pointing position of the telescopes was set as a wobble position around the source (the ON region center), with an offset fixed at 0.5 degrees.  

In the region of the LHAASO J1825-1326, two sources were modeled: CTA I - near the pulsar PSR J1826-1256 and CTA II - near the pulsar PSR J1826-1334. Tables \ref{tab:3} and \ref{tab:4} present each counterpart's celestial coordinates and angular separation used in the likelihood fit for CTA I and CTA II, respectively. The Table~\ref{tab:5} displays the source spectral model parameters found for each case, such as the celestial coordinates and size established. Regarding the LHAASO J1839-0545, two sources were modeled: CTA III - near the pulsar PSR J1837-0604 and CTA IV - near the PSR J1838-0537. Tables \ref{tab:6} and \ref{tab:7} show each counterpart's celestial coordinates and angular separation considered in the likelihood fit for CTA III and CTA IV, respectively. The spectral model parameters determined for each case, such as the source's celestial coordinates and size, are displayed in Table \ref{tab:8}.

\begin{table}[htbp]
\begin{minipage}{\textwidth}
\centering
\begin{tabular}{lccc}
\toprule
Counterpart & R.A. (deg.)& Dec. (deg.)& Sep. (deg.)\\
\hline     
\midrule
    4FGL J1824.1-1304 \cite{ballet2023fermi, Abdollahi2022}& 276.03 &    -13.07 &      0.5 \\
    4FGL J1826.1-1256 \cite{ballet2023fermi, Abdollahi2022}& 276.54 &    -12.94 &      0.1 \\
    4FGL J1828.1-1312 \cite{ballet2023fermi, Abdollahi2022}& 277.03 &    -13.20 &      0.5 \\
    HAWC J1826-128 \cite{Albert2021}&  276.50 &    -12.86 &      0.2 \\
    HESS J1826-130 \cite{Abdalla2018}$^{a}$&  276.51 &    -13.02 &      0.0 \\ 
    LHAASO J1825-1326 \cite{Cao2021a}&  276.45 & -13.45 &   0.4 \\
\bottomrule
\end{tabular}
\caption{Celestial coordinates (R.A., Dec.) and the angular separation of the counterparts from the CTA I source. \label{tab:3}}
\footnotesize{$^{a}$ The HESS instrument data can be found at: \href{https://www.mpi-hd.mpg.de/hfm/HESS/hgps/}{https://www.mpi-hd.mpg.de/hfm/HESS/hgps/}.}
\end{minipage}
\end{table}

\begin{table}[htbp]
\begin{minipage}{\textwidth}
\centering
\begin{tabular}{lccc}
\toprule
Counterpart & R.A. (deg.)& Dec. (deg.)& Sep. (deg.)\\
\hline     
\midrule
    4FGL J1824.4-1350e \cite{Abdollahi2022, ballet2023fermi}&  276.11 &     -13.84 &       0.3  \\  
    HAWC J1825-138 \cite{Albert2021}&  276.38 &     -13.86 &       0.1 \\
    HESS J1825-137 \cite{Abdalla2019}$^{a}$
    &  276.45 &    -13.78 &    0.0  \\
    LHAASO J1825-1326 \cite{Cao2021a}&  276.45 & -13.45 & 0.3\\
\bottomrule
\end{tabular}
\caption{
Celestial coordinates (R.A., Dec.) and the angular separation of the counterparts from the CTA II source.
} \label{tab:4}
\footnotesize{$^{a}$ The HESS instrument data can be found at: \href{https://cdsarc.cds.unistra.fr/ftp/J/A+A/621/A116/}{https://cdsarc.cds.unistra.fr/ftp/J/A+A/621/A116/}.}
\end{minipage}
\end{table}

\begingroup
\renewcommand*{\thefootnote}{\alph{footnote}}  
    \begin{table}[htbp]
    \centering
        \begin{minipage}{\textwidth}
        \centering
            \begin{tabular}{lll}
            \toprule
             source & CTA I & CTA II\\
             \hline     
\midrule
                 index $\rm{\Gamma}$ & 1.56 $\pm$ 0.07  &  2.47 $\pm$ 0.01 \\
                 amplitude $\rm{\phi_0\ (cm^{-2}\ s^{-1}\ TeV^{-1})}$&  (3.09 $\pm$ 0.22) $\times 10^{-14}$ &  (5.86 $\pm$ 0.14) $\times 10^{-14}$\\
                 cutoff energy $\rm{E_{cut}\ (TeV)}$ &75 $\pm$  11 & 283 $\pm$ 69\\
                 location R.A., Dec. (deg.)& $\rm{276.51, -13.02}$  \cite{Abdalla2019}& $\rm{276.45, -13.78}$  \cite{Abdalla2018}\\
                 size (deg.) &  0.46  $\pm$ 0.03 \cite{Abdalla2018}&  0.15  $\pm$ 0.02 \cite{Abdalla2018} \\
            \bottomrule
            \end{tabular}
            \caption[]{
            Specifications of the spectral model parameters, such as the celestial coordinates and size, for the sources CTA I and CTA II around the LHAASO J1825-1326 source.
            }    
            \label{tab:5}
        \end{minipage}
    \end{table}
\endgroup

\begin{table}[htbp]
    \begin{minipage}{\textwidth}
    \centering
        \begin{tabular}{lccc}
        \toprule
        Counterpart & R.A. (deg.)& Dec. (deg.)& Sep. (deg.)\\
        \hline     
\midrule
            2HWC J1837-065 \cite{Abeysekara2017}&  279.36 &  -6.58 &   0.4\\
            HESS J1837-069 \cite{Abdalla2018}&  279.37 &     -6.96 &      0.0\\
            LHAASO J1839-0545 \cite{Cao2021a}&  279.95 &  -5.75 &   1.3\\
        \bottomrule
        \end{tabular}
    \caption{Celestial coordinates (R.A., Dec.) and the angular separation of the counterparts from the CTA III source.} \label{tab:6}
    \end{minipage}
\end{table}

\begin{table}[htbp]
    \begin{minipage}{\textwidth}
    \centering
        \begin{tabular}{lccc}
        \toprule
        Counterpart & R.A. (deg.)& Dec. (deg.)& Sep. (deg.)\\
       \hline     
\midrule
            4FGL J1840.9-0532e  \cite{Abdollahi2022, ballet2023fermi}&  280.23 &      -5.55 &       0.1\\
            HESS J1841-055 \cite{Abdalla2018}&  280.22 &      -5.64 &       0.0\\
            LHAASO J1839-0545 \cite{Cao2021a}&  279.95 &      -5.75 &       0.3\\
        \bottomrule
        \end{tabular}
    \caption{Celestial coordinates (R.A., Dec.) and the angular separation of the counterparts from the CTA IV source.} \label{tab:7}
    \end{minipage}
    \end{table}

\begingroup
\renewcommand*{\thefootnote}{\alph{footnote}}  
    \begin{table}[htbp]
    \centering
        \begin{minipage}{\textwidth}
        \centering
            \begin{tabular}{lll}
            \toprule
             source & CTA III & CTA IV\\
            \hline     
\midrule
                 index $\rm{\Gamma}$ & 2.35 $\pm$ 0.09  & 2.07 $\pm$  0.08\\
                 amplitude $\rm{\phi_0\ (cm^{-2}\ s^{-1}\ TeV^{-1})}$&   (9.21 $\pm$ 2.37) $\times 10^{-14}$  & (16.00 $\pm$ 3.62) $\times 10^{-14}$\\
                 cutoff energy $\rm{E_{cut}\ (TeV)}$ & 13 $\pm$ 7  & 7 $\pm$ 2\\
                 location R.A., Dec. (deg.)& $\rm{279.37, -6.96}$  \cite{Abdalla2018}& $\rm{280.22, -5.64}$ \cite{Abdalla2018}\\
                 size (deg.) & 0.36 $\pm$ 0.03 \cite{Abdalla2018}&  0.41 $\pm$ 0.03 \cite{Abdalla2018}\\
            \bottomrule
            \end{tabular}
        \caption[]{Specifications of the spectral model parameters, such as the celestial coordinates and size, for the sources CTA III and CTA IV around the LHAASO J1839-0545 source.
        }    
        \label{tab:8}
        \end{minipage}
    \end{table}
\endgroup

To compare the performance of the two CTAO observation arrays (CTAO South and CTAO North) by observing the sources modeled, we utilized the IRFs: The CTAO North with 4 LSTs (Large Sized Telescopes) and 9 MSTs (Medium Sized Telescopes); CTAO North-LSTs with 4 LSTs; CTAO North-MSTs with 9 MSTs; CTAO South with 14 MSTs and 37 SSTs (Small Sized Telescopes); CTAO South-MSTs
with 14 MSTs and CTAO South-SSTs with 37 SSTs. All IRFs are optimized for 50 hours of observation time.  As is recommended, in the CTAO IRFs documentation \cite{CTAIRFS2021}, for general studies, all IRFs are azimuth-averaged pointing. To account for variation in the point spread function (PSF) leakage, the integration radius (see the source size in the Tables \ref{tab:5} and~\ref{tab:8}) was considered to be energy-dependent. A constant fraction of $68\%$ of flux enclosure was fixed to correct the energy-dependent region size and the background estimation, and it imposed a minimal number of expected signal counts of 5 per bin and a minimal significance of 3 $\sigma$ per bin. The gamma-rays excess must also be greater than $10\%$ of the background count.

\begin{figure}[h!]
\centering
\subfloat[CTA I - PSR J1826-1256 region.]
{\includegraphics[angle=0,width=0.5\textwidth]{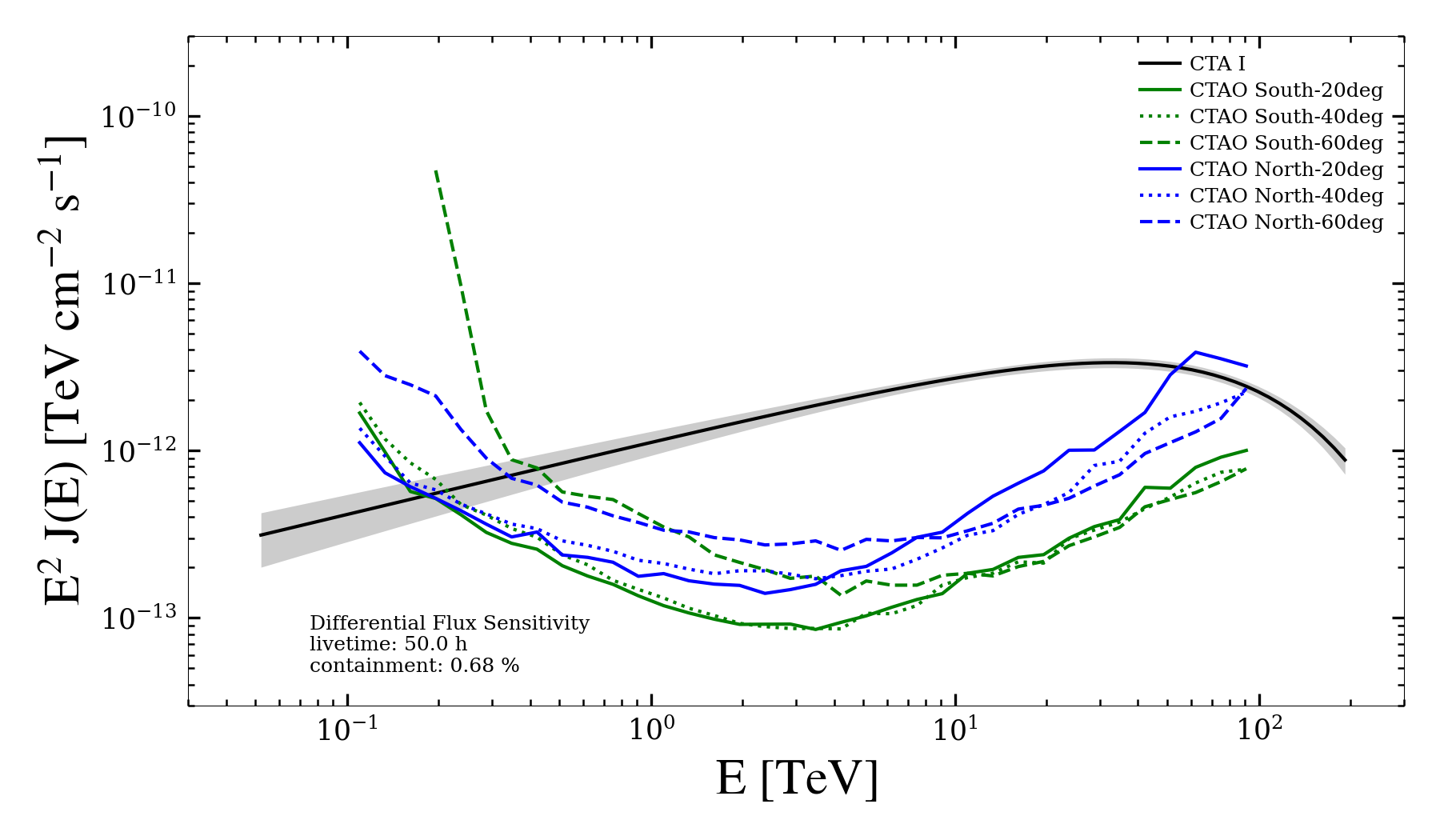}}
\subfloat[CTA II - PSR J1826-1334 region.]
{\includegraphics[angle=0,width=0.5\textwidth]{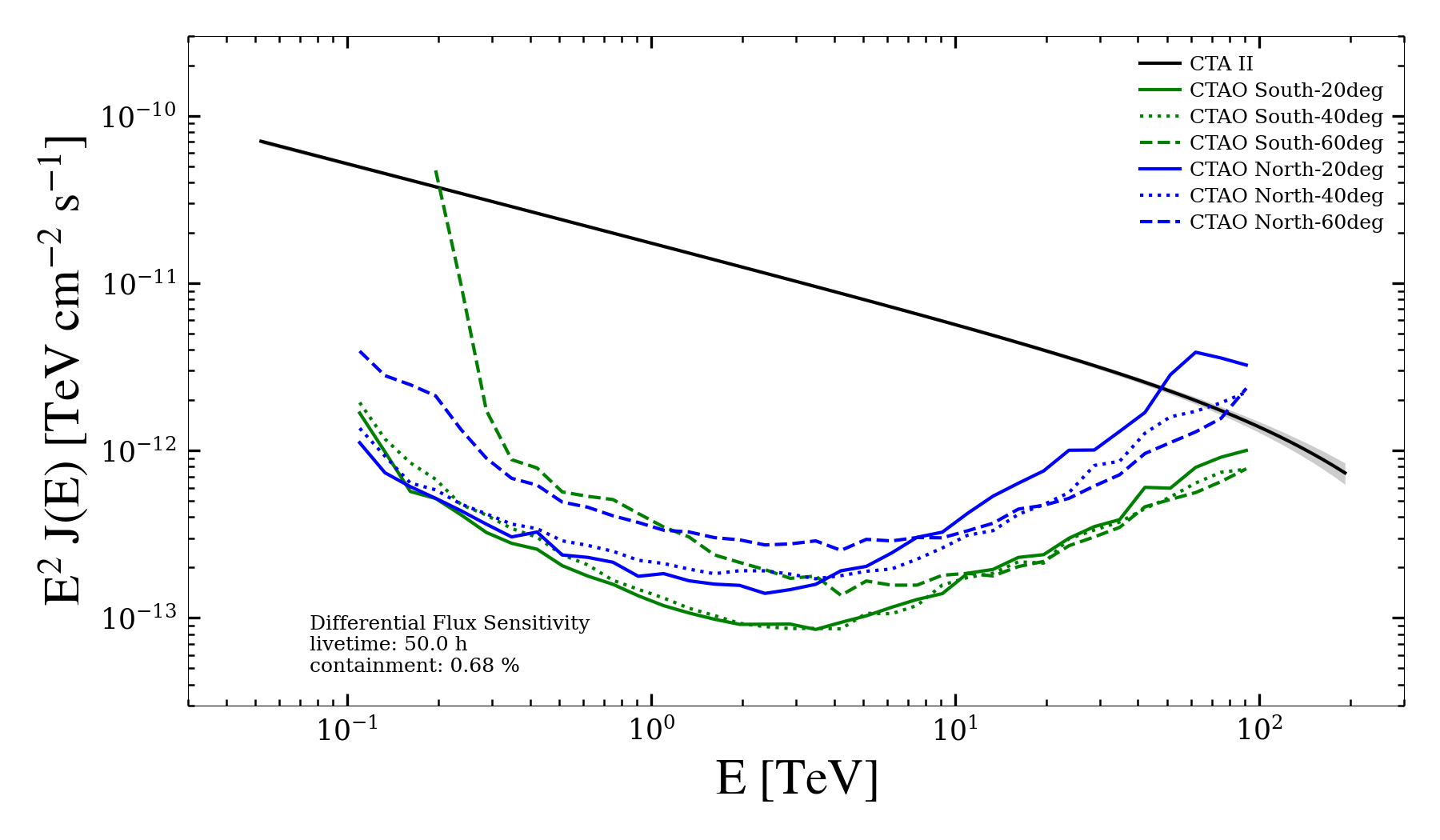}}\\
\subfloat[CTA III - PSR J1837-0604 region.]
{\includegraphics[angle=0,width=0.5\textwidth]{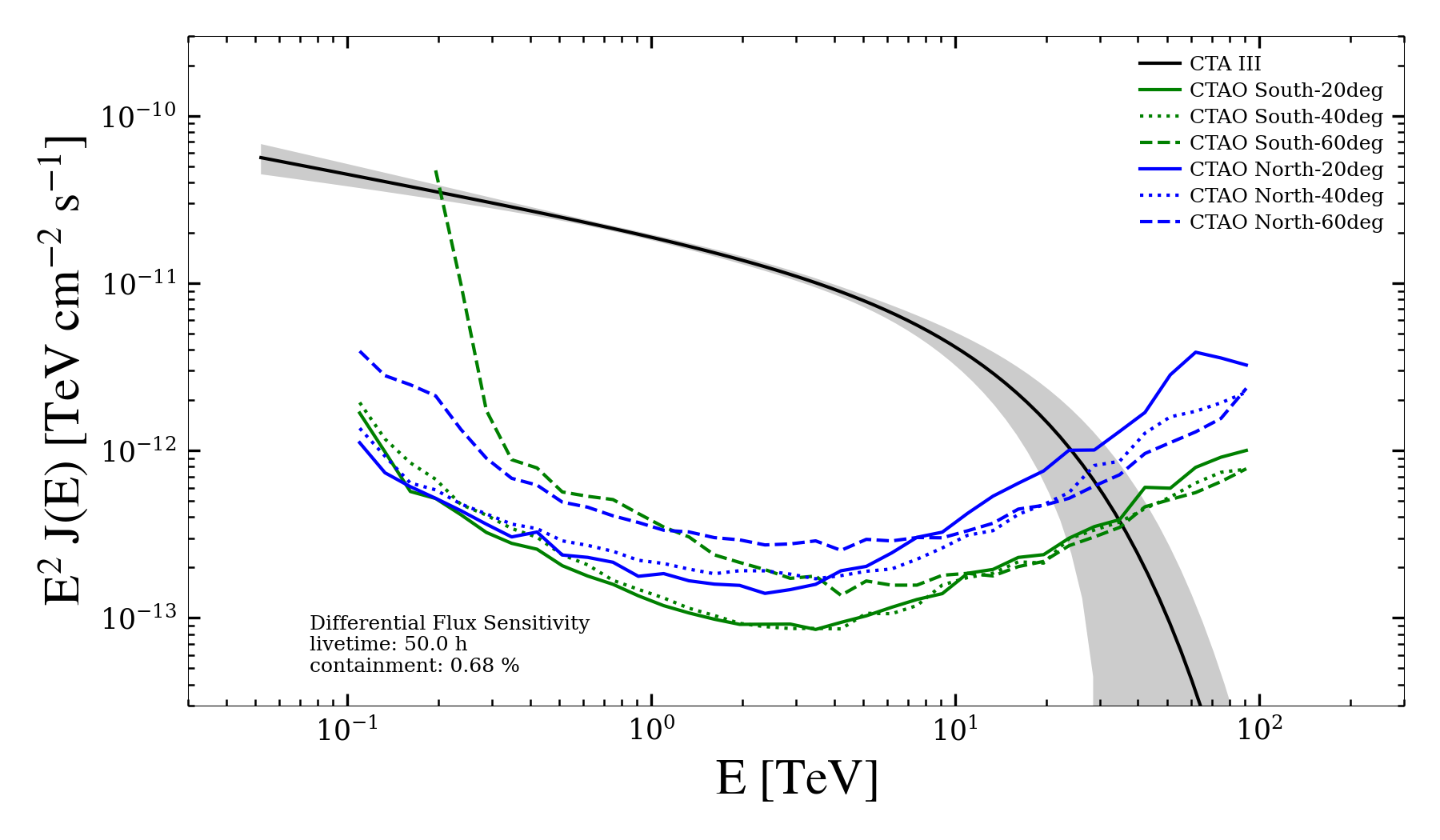}}
\subfloat[CTA IV -  PSR J1838-0537 region.]
{\includegraphics[angle=0,width=0.5\textwidth]{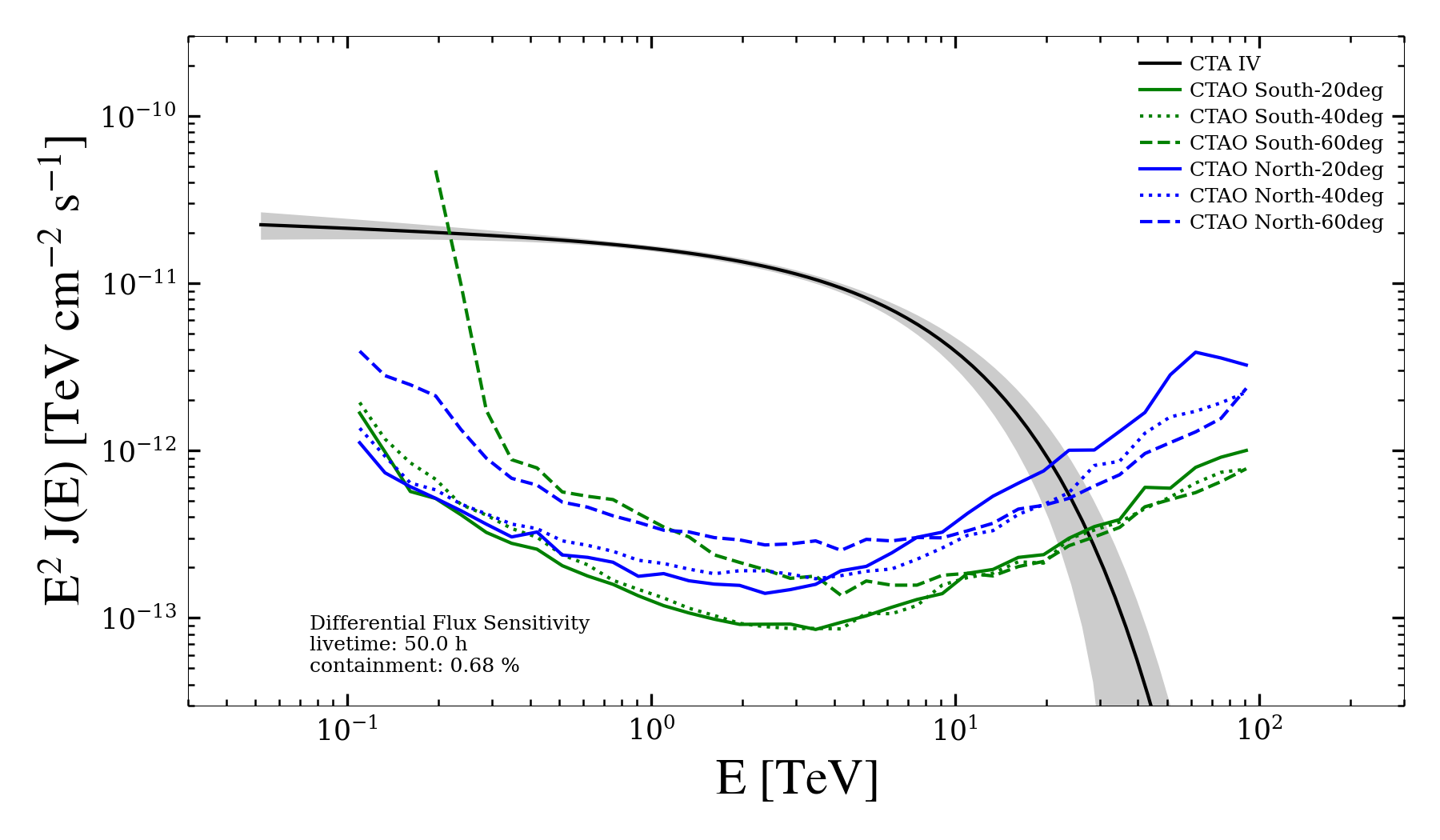}}
\caption{Source spectral energy distribution and CTAO differential sensitivity curves of the Northern and Southern CTAO arrays (for 50 hours of observations).}
\label{fig:systems}
\end{figure}

\begin{figure}[h!]
\centering
\subfloat[CTA I - PSR J1826-1256 region.]
{\includegraphics[angle=0,width=0.5\textwidth]{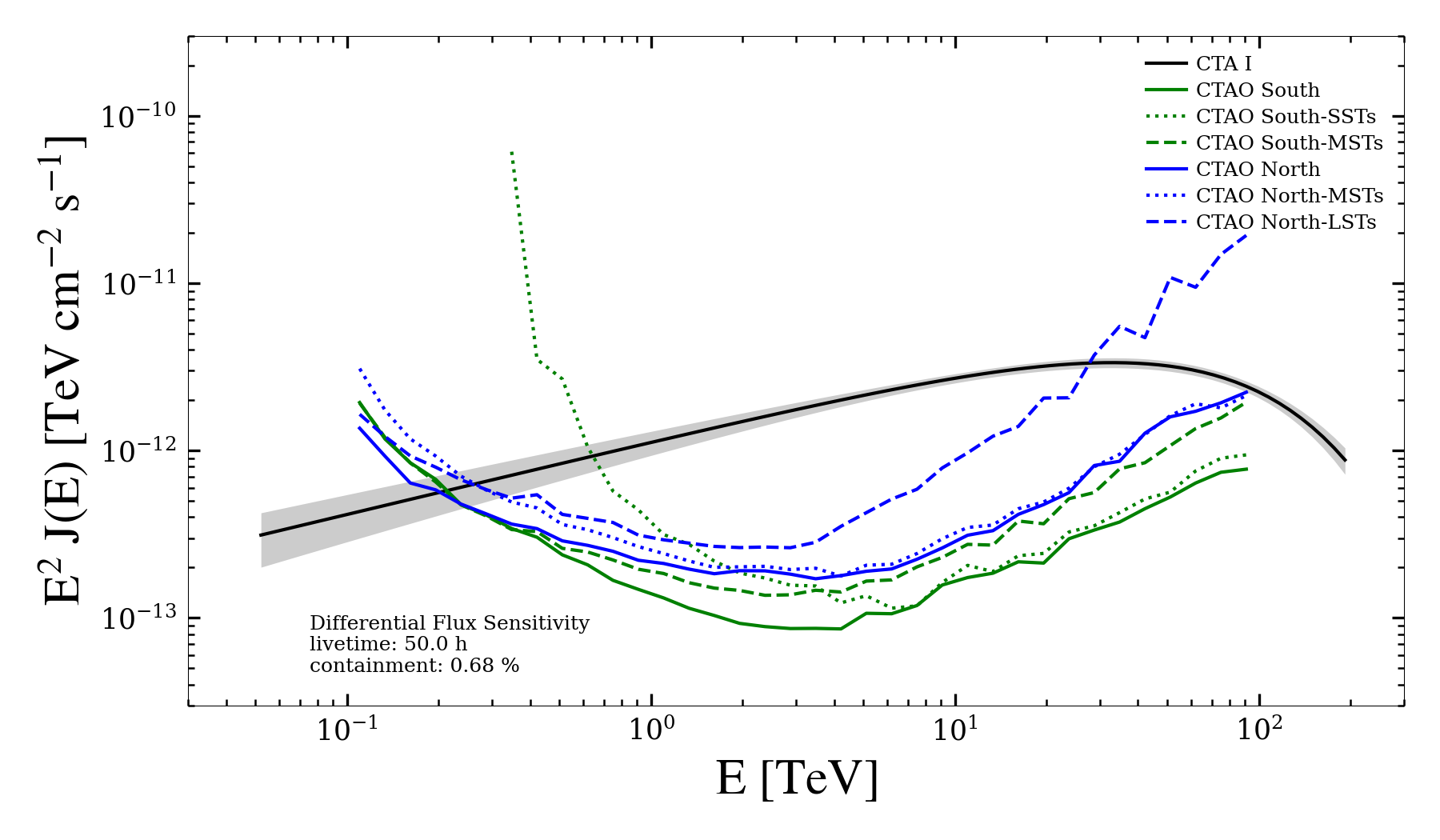}}
\subfloat[CTA II - PSR J1826-1334 region.]
{\includegraphics[angle=0,width=0.5\textwidth]{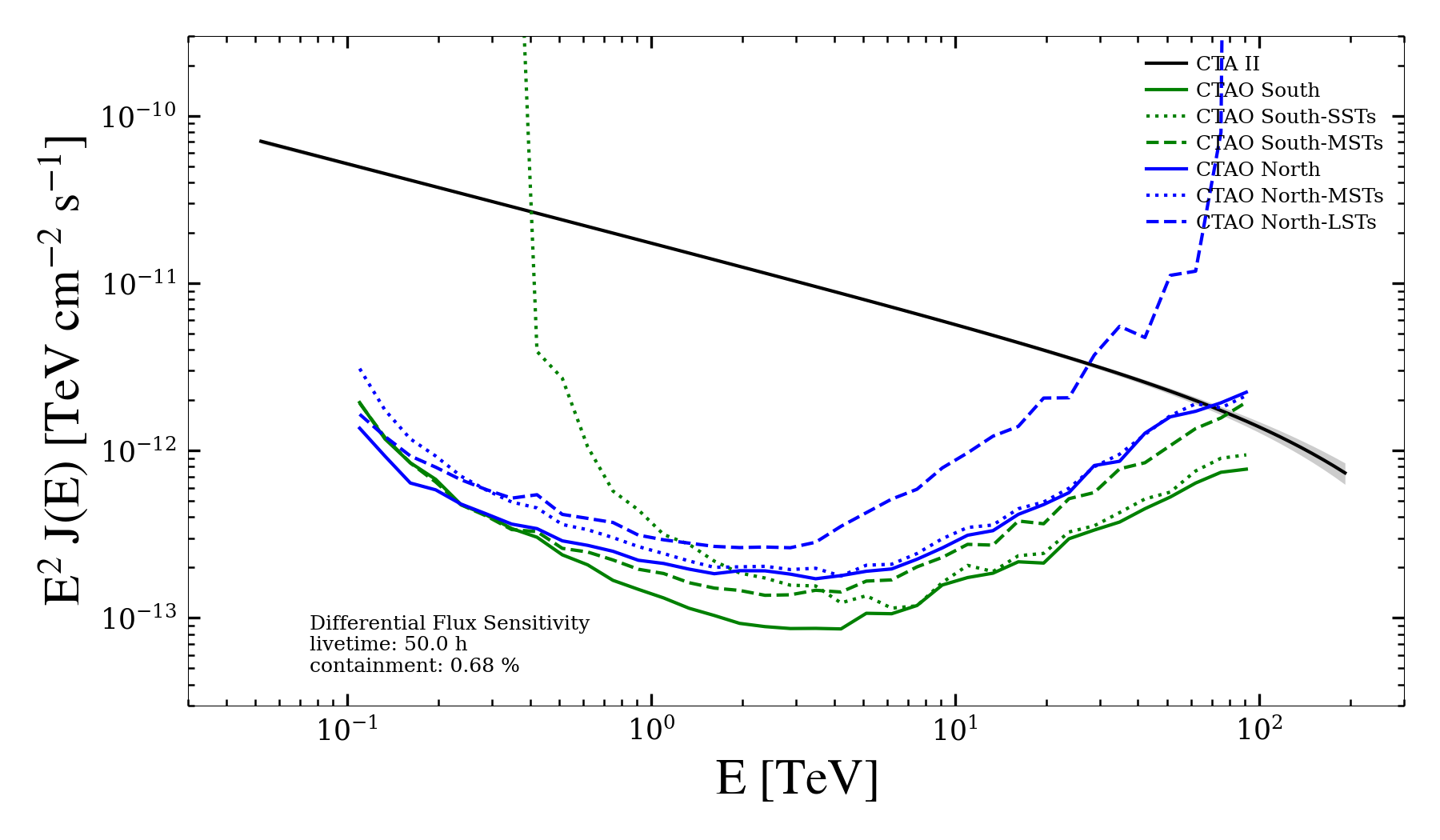}}\\
\subfloat[CTA III - PSR J1837-0604 region.]
{\includegraphics[angle=0,width=0.5\textwidth]{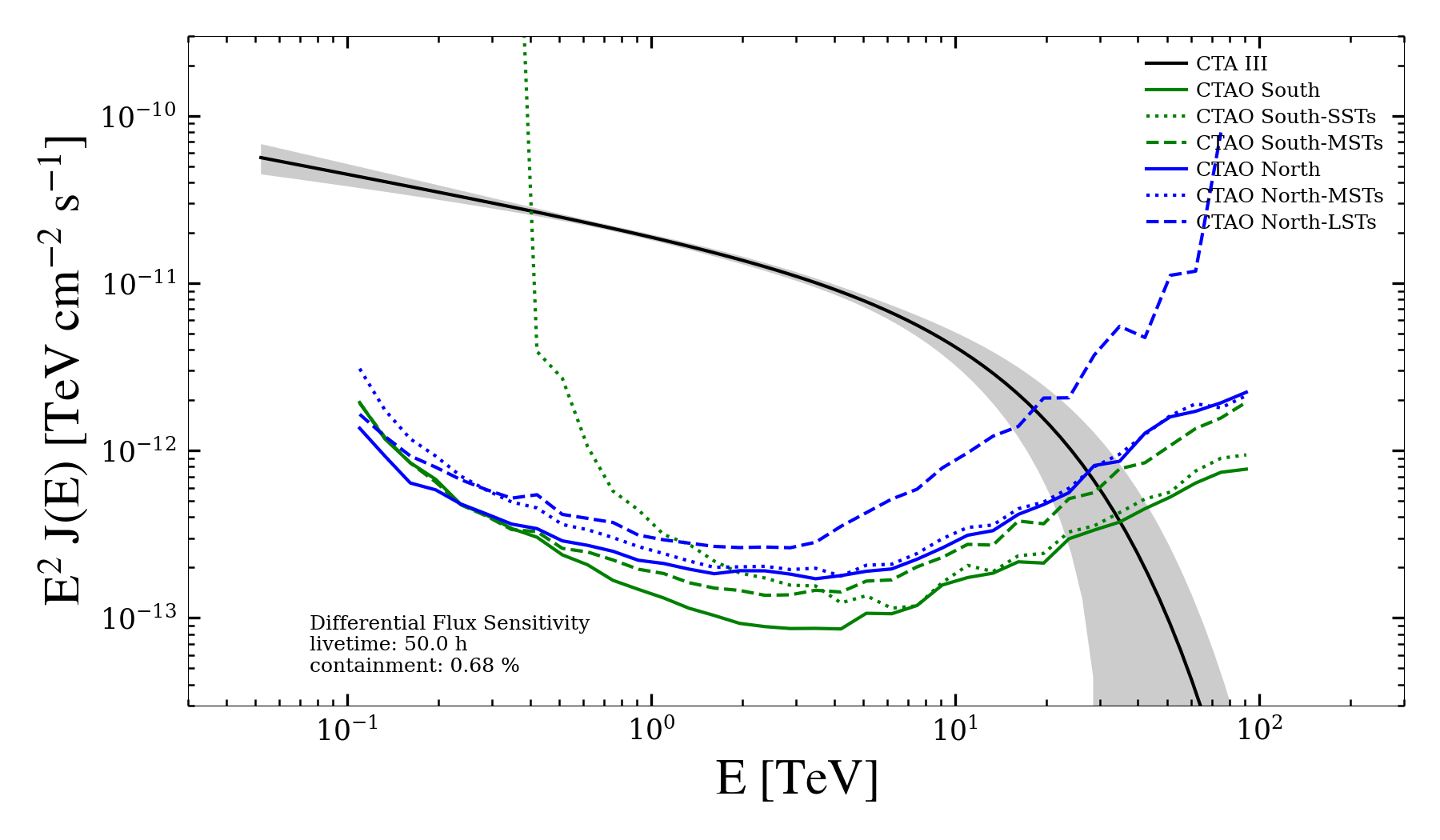}}
\subfloat[CTA IV - PSR J1838-0537 region.]
{\includegraphics[angle=0,width=0.5\textwidth]{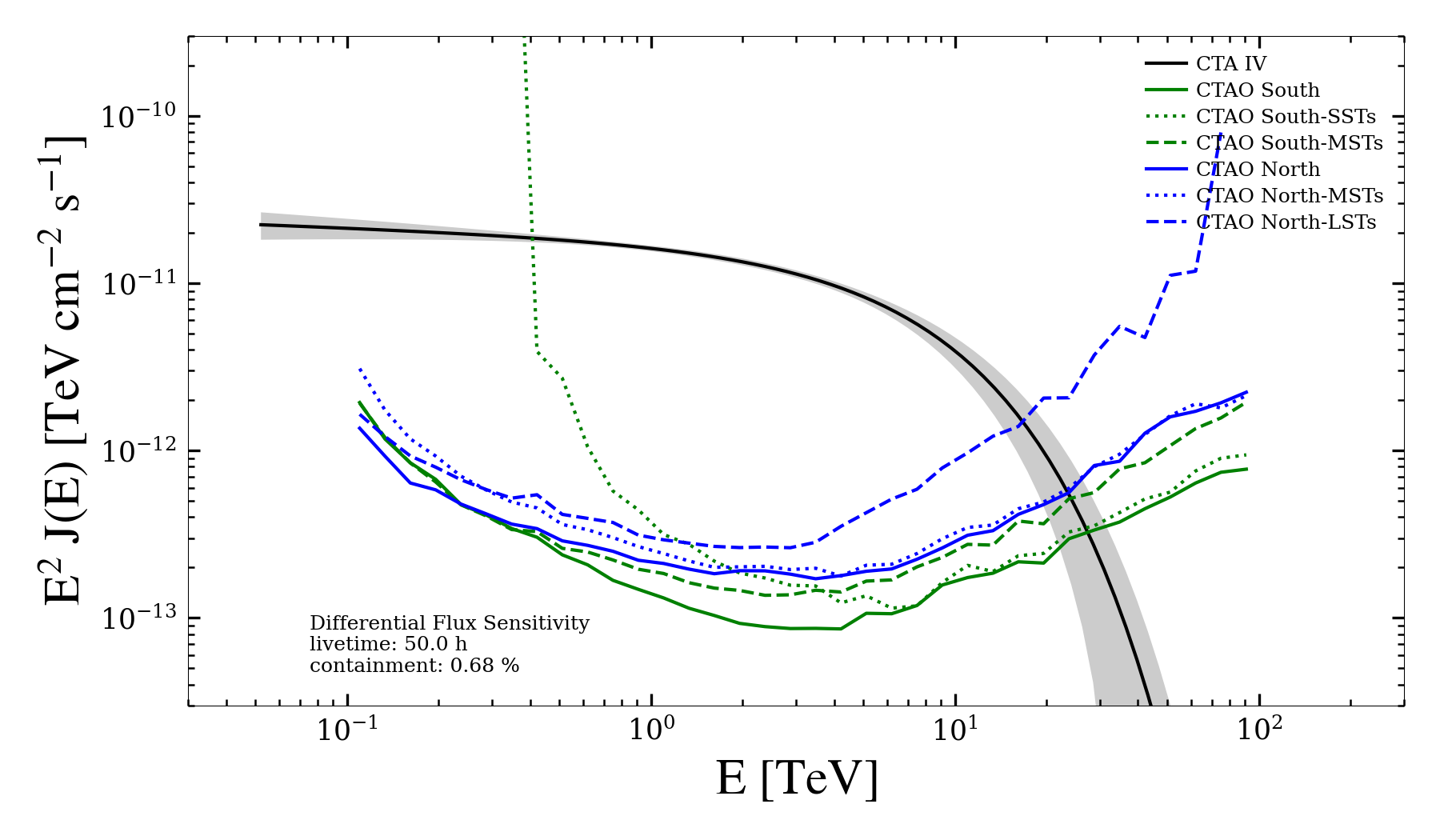}}
\caption{Source spectral energy distribution and CTAO differential sensitivity curves of the Northern and Southern CTAO subarrays (for 50 hours of observations).}
\label{fig:sub-systems}
\end{figure}

We generated sensitivity curves for the full array's IRFs optimized for the three zenith angles (20, 40, and 60 degrees) (see Figure~\ref{fig:systems}).  Regarding the CTA I source, in all considered array configurations, the sensitivity
was insufficient to quantify flux in the lower energy ranges of the
spectrum (see Figure~\ref{fig:systems}-(a)). At the beginning of the spectrum, for all sources analyzed, the sensitivity of the CTAO South optimized for the zenith angle of 60 degrees was not sufficient. The necessary sensitivity is reached only for energies $\gtrsim 0.4\rm~TeV$ for the source CTA I (see Figure~\ref{fig:systems}-(a)) and energies $\gtrsim 0.2\rm~TeV$ for the sources CTA II, CTA III, and CTA IV (see Figures~\ref{fig:systems}-(b),~\ref{fig:systems}-(c), and~\ref{fig:systems}-(d), respectively).  At the end of the spectrum, all configurations faced challenges in reaching the minimum sensitivity for the sources CTA III (see Figure~\ref{fig:systems}-(c)) and CTA IV  (see Figure~\ref{fig:systems}-(d)).  The CTAO South configurations reached the necessary sensitivity only for energies $\lesssim 40\ \rm~TeV$ and the CTAO North configurations  for energies $\lesssim 30\ \rm~TeV$. The sensitivity of the CTAO North, optimized for a zenith angle of 20 degrees, is sufficient only for energies $\lesssim 60\rm~TeV$ in the cases of the sources CTA I (see Figure~\ref{fig:systems}-(a)) and CTA II (see Figure~\ref{fig:systems}-(b)). We also compared the performance of the different arrays and subarrays, considering IRFs optimized for 50 hours of observation and a zenith angle of 40 degrees (see Figure \ref{fig:sub-systems}). In all cases analyzed, the CTAO South-SSTs sensitivity was insufficient at the beginning of the spectrum. It reached the necessary sensitivity only for energies $\gtrsim 0.6\rm~TeV$ for the source CTA I (see Figure~\ref{fig:sub-systems}-(a)) and energies $\gtrsim 0.4\rm~TeV$ for the sources CTA II, CTA III, and CTA IV (see Figures~\ref{fig:sub-systems}-(b),~\ref{fig:sub-systems}-(c), and~\ref{fig:sub-systems}-(d), respectively). The CTAO North-LSTs sensitivity curves show the worst performance at the end of the spectrum. It reached the necessary sensitivity only for energies $\lesssim 30\rm~TeV$ in the case of the source CTA I (see Figure~\ref{fig:sub-systems}-(a)), for energies $\lesssim 40\rm~TeV$ in the case of the source CTA II (see Figure~\ref{fig:sub-systems}-(b)), and for energies $\lesssim 20\rm~TeV$ in the cases of the sources CTA III and CTA IV (see Figures~\ref{fig:sub-systems}-(c) and~\ref{fig:sub-systems}-(d), respectively). After analyzing the sensitivity curves for both arrays and subarrays utilizing IRFs for 50 hours (see Figures \ref{fig:systems} and \ref{fig:sub-systems}), it was evident that the CTAO South full array's IRFs exhibited better performance when observing the modeled sources.

\begin{figure}[htbp]
\centering
\includegraphics[width=1\textwidth]{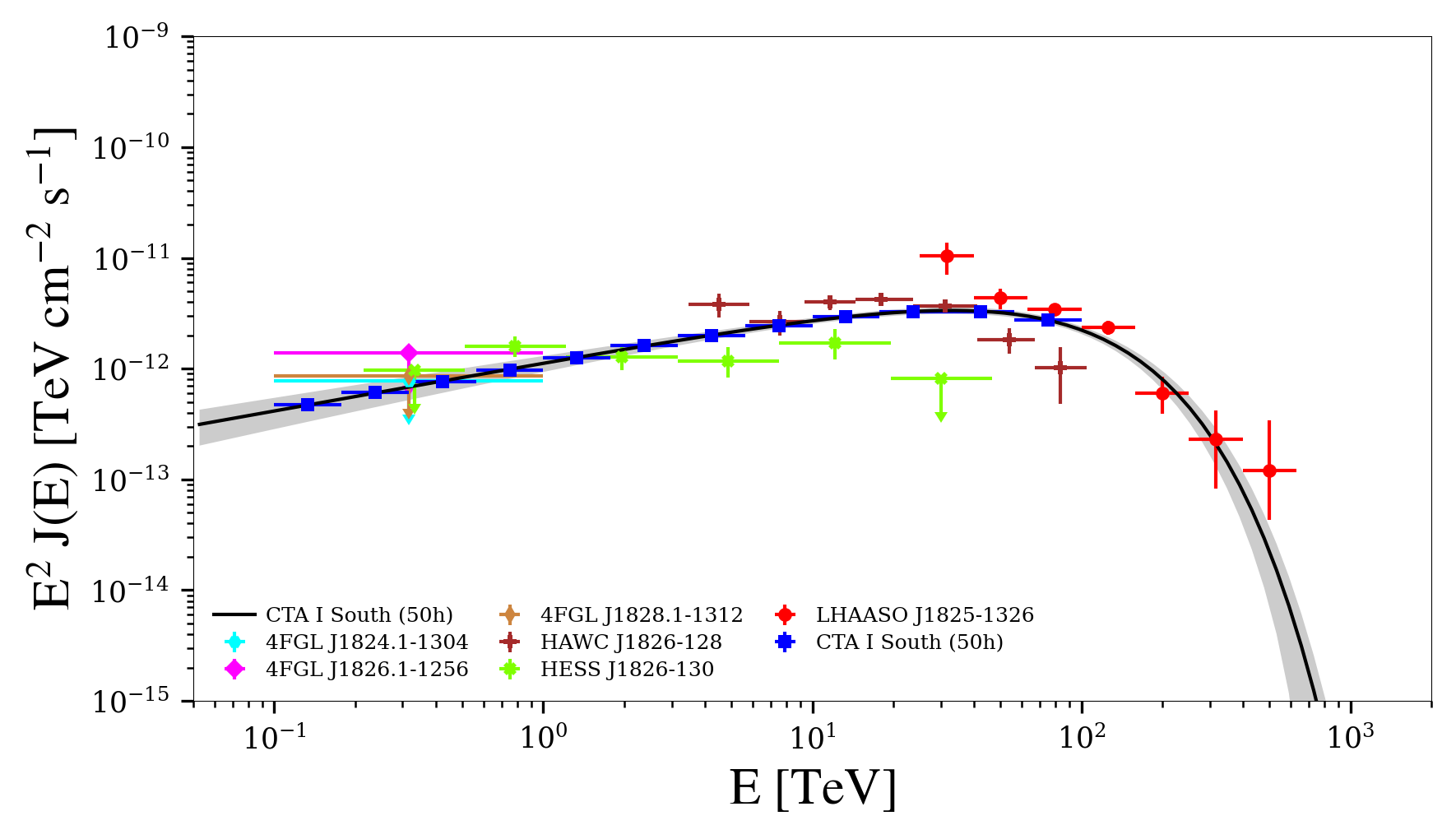}
\caption{CTA I (PSR J1826-1256 region) spectral energy distribution and previous observations of the counterparts included in the likelihood analysis, as Table \ref{tab:3} displays. The solid line shows the simulated spectral fits, the parameters of which are listed in Table \ref{tab:9}.}
\label{fig:ctaI}
\end{figure}

\begin{figure}[htbp]
\centering
\includegraphics[width=1\textwidth]{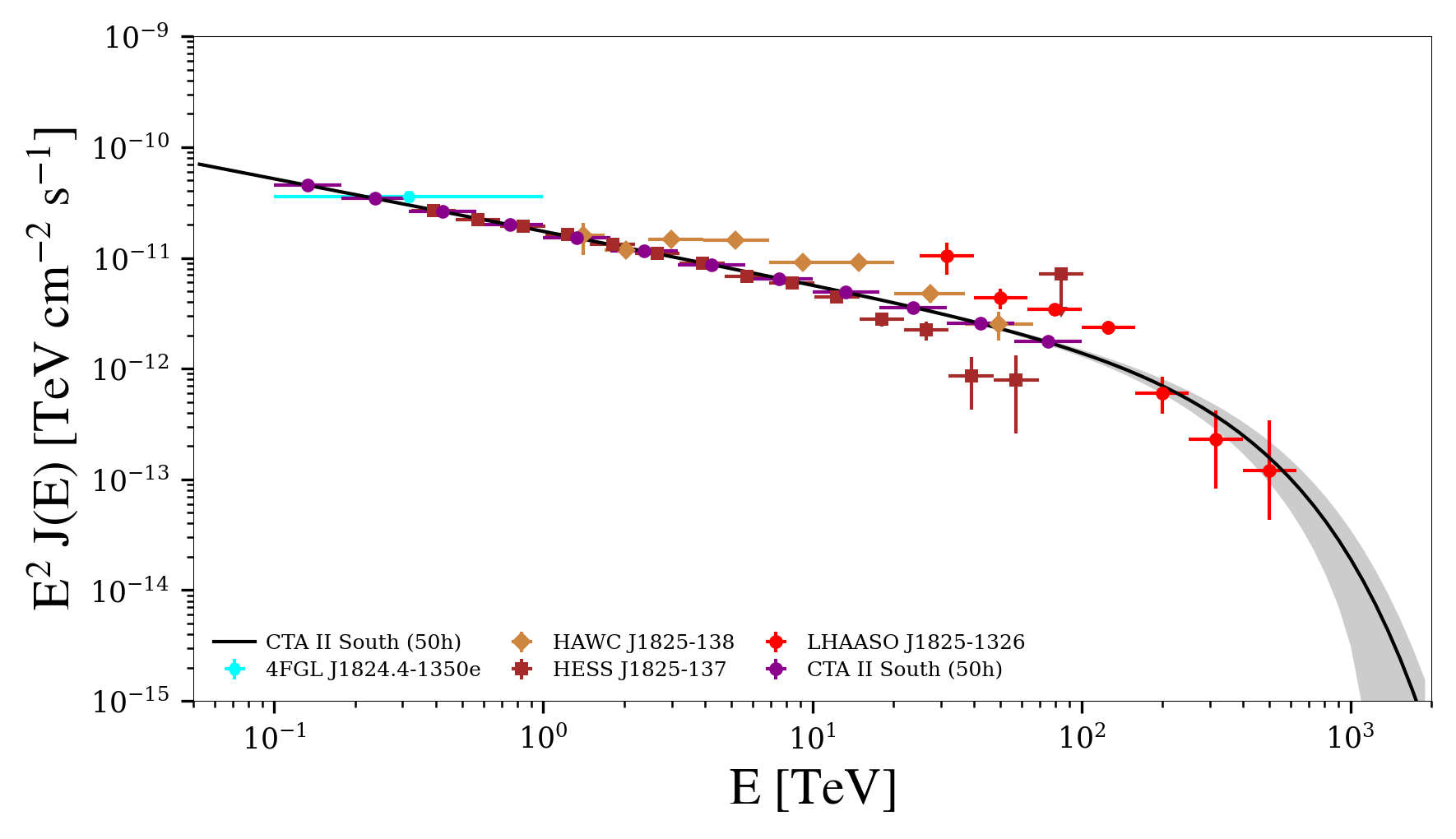}
\caption{CTA II (PSR J1826-1334 region) spectral energy distribution and previous observations of the counterparts included in the likelihood fit analysis, as Table \ref{tab:4} displays. The solid line shows the simulated spectral fits, the parameters of which are listed in Table \ref{tab:9}.}
\label{fig:ctaII}
\end{figure}

\begin{figure}[htbp]
\centering
\includegraphics[width=1\textwidth]{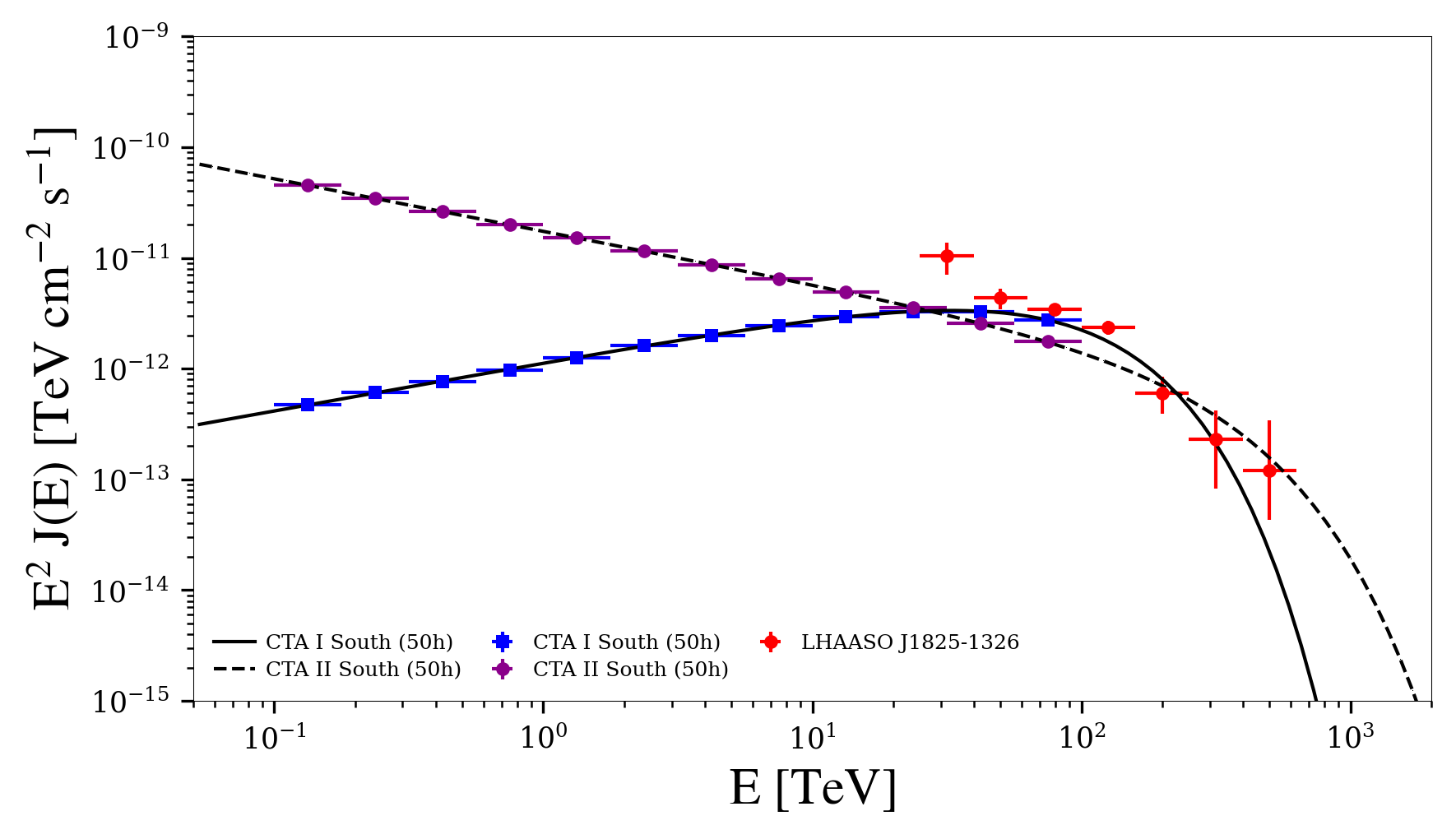}
\caption{Spectral energy distribution of the sources CTA I and CTA II, along with the LHAASO J1825-1326 source. \label{fig:3}}
\end{figure}

\begin{figure}[htbp]
\centering
\includegraphics[width=1\textwidth]{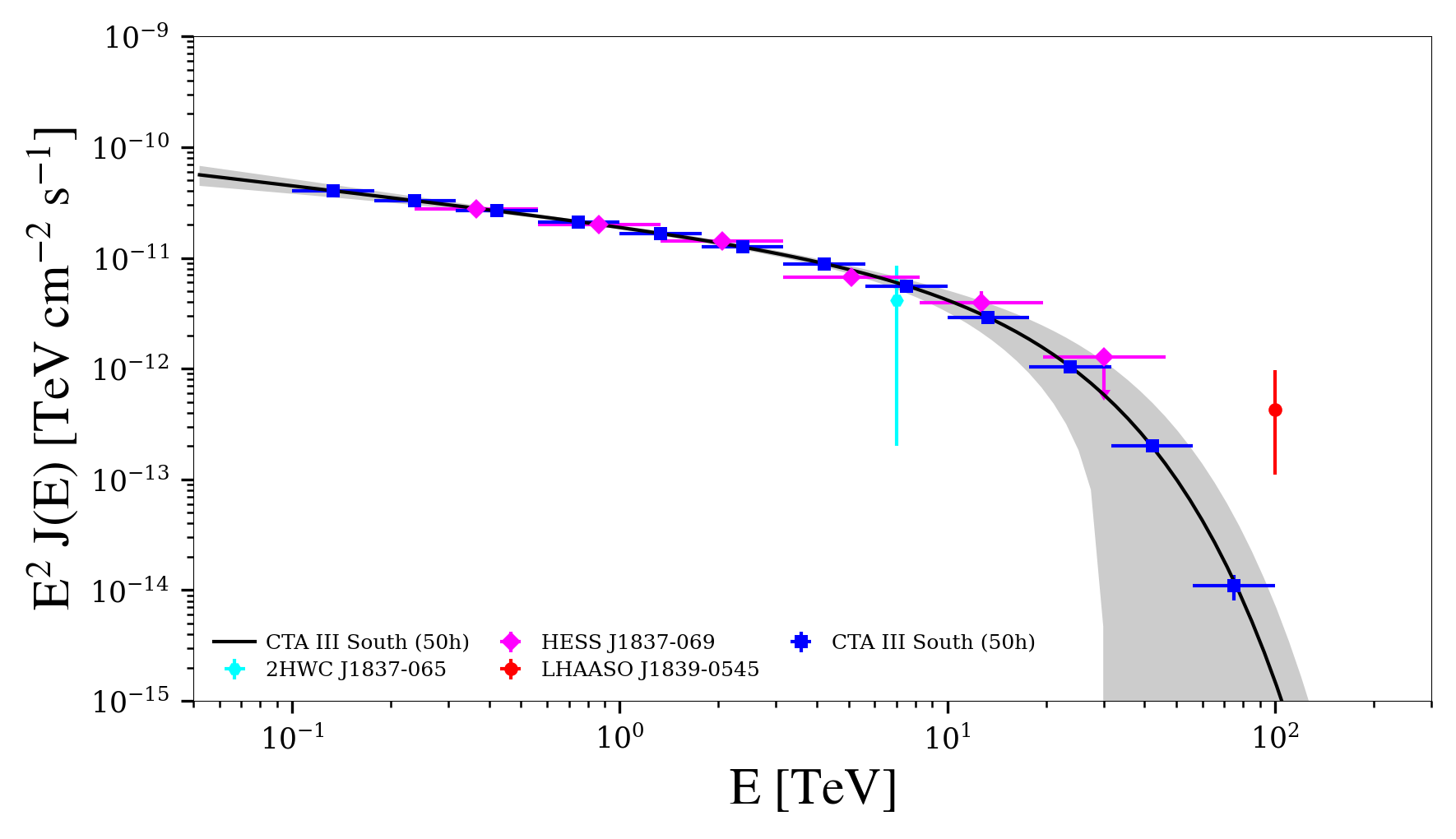}
\caption{CTA III (PSR J1837-0604 region) spectral energy distribution and previous observations of the counterparts included in the likelihood analysis, as Table \ref{tab:6} displays. The solid line shows the simulated spectral fits, the parameters of which are listed in Table \ref{tab:9}.}
\label{fig:ctaIII}
\end{figure}

\begin{figure}[htbp]
\centering
\includegraphics[width=1\textwidth]{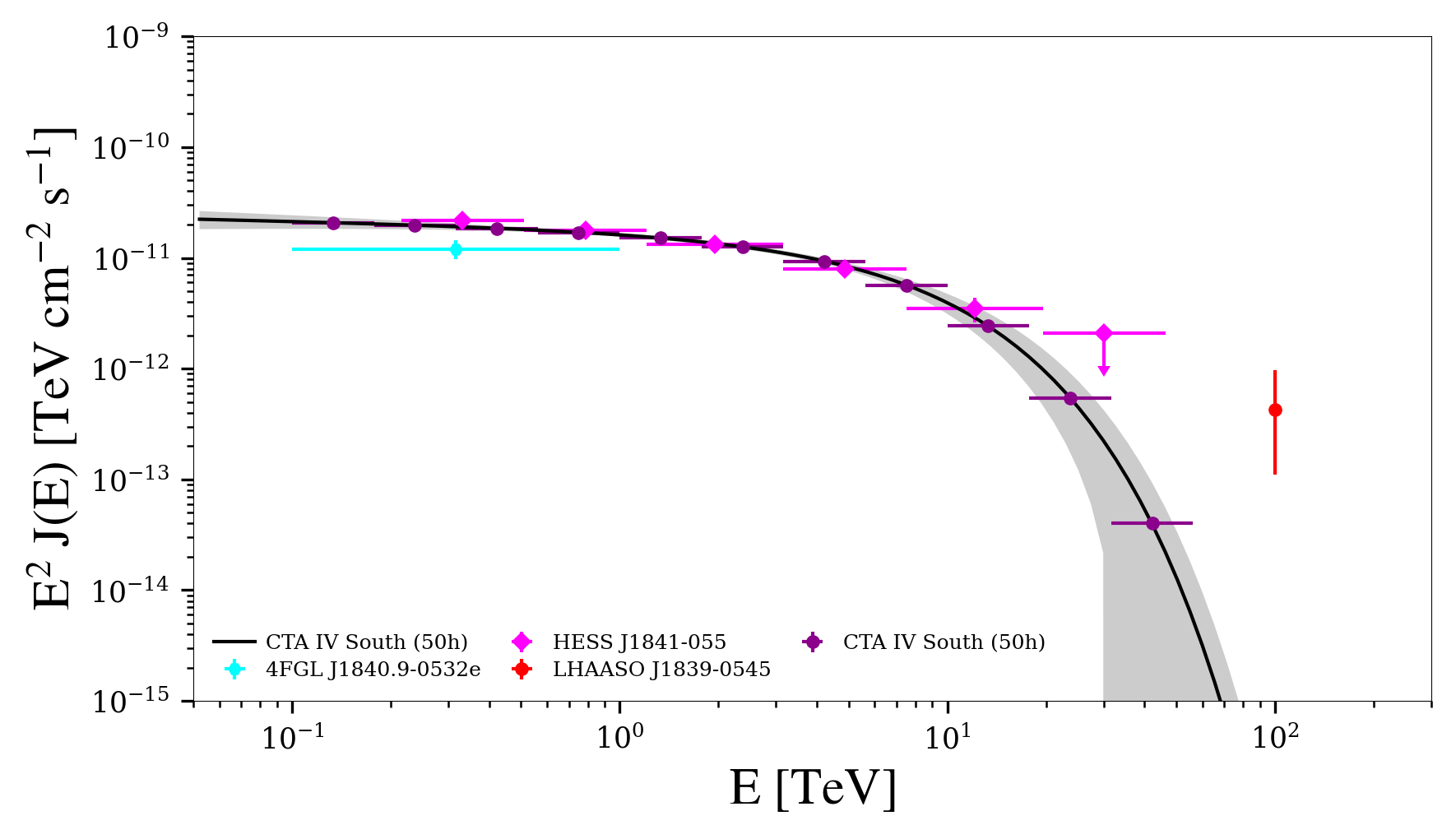}
\caption{CTA IV (PSR J1838-0537 region) spectral energy distribution and previous observations of the counterparts included in the likelihood analysis, as Table \ref{tab:7} displays. The solid line shows the simulated spectral fits, the parameters of which are listed in Table \ref{tab:9}.}
\label{fig:ctaIV}
\end{figure}

\begin{figure}[htbp]
\centering
\includegraphics[width=1\textwidth]{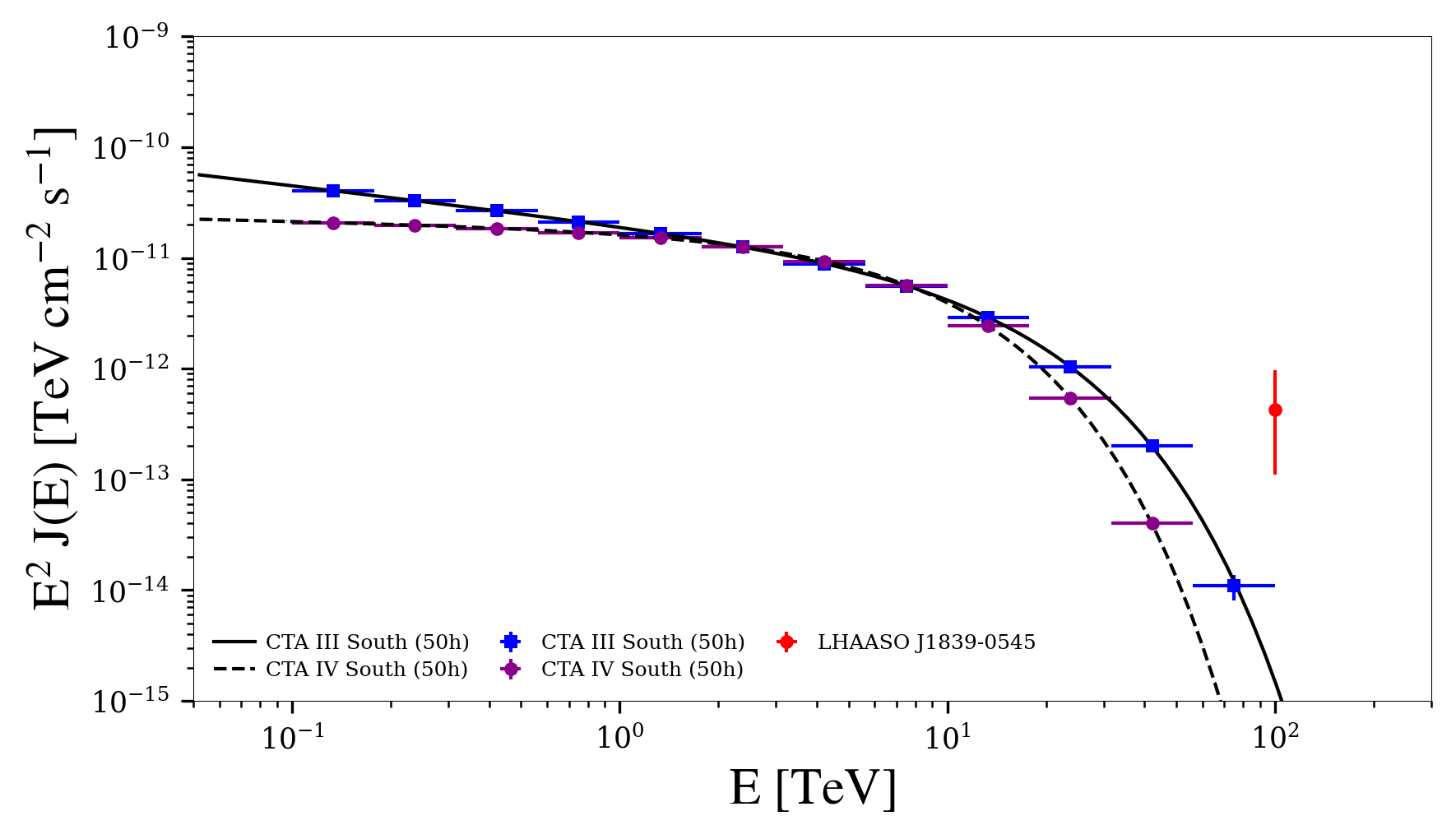}
\caption{Spectral energy distribution of CTA III and CTA IV, along with the LHAASO J1839-0545 source.\label{fig:6}}
\end{figure}

To estimate the flux points, we simulated a thousand CTAO observations of each source modeled, using the CTAO South full array's IRFs optimized for the zenith angle of 40 degrees. We generated gamma rays in the energy range of 100 GeV to 100 TeV, which is narrower than the designed energy range of CTAO in full-array configurations. Figure \ref{fig:ctaI} depicts the results for CTA I as well as the earlier observations of potential counterparts. A 0.076 degrees integration radius was considered, and the maximum detectable cutoff energy in the CTAO simulated spectrum is up to $\sim$ 75 TeV. As depicted in Figure \ref{fig:ctaI}, in this scenario, the CTAO observation can detect an exponential decrease in the spectrum above 100 TeV, and the contribution from a single region was sufficient to explain the LHAASO observation.  In other words, the simulated spectra display an exponentially decreasing trend that closely follows the entire LHAASO spectrum. This alignment suggests that the observation may originate from the vicinity of PSR J1826-1256.
The simulation results for CTA II and previous observations of potential counterparts are described in Figure \ref{fig:ctaII}. In this scenario, an integration radius of 0.231 degrees was considered, and the maximum detectable cutoff energy in the CTAO simulated spectrum is $\sim$ 299 TeV. While the CTAO observation can detect an exponential decrease in the spectrum above 100 TeV, the
LHAASO observation includes the contribution of the two pulsar regions at lower energies. The two simulations of the CTAO observations are illustrated in Figure \ref{fig:3} alongside the LHAASO observation. The simulation results for the observation CTA III and previous observations of potential counterparts are presented in Figure \ref{fig:ctaIII}. An integration radius of 0.178 degrees was employed in this scenario. The simulation results for CTA IV and previous observations of potential counterparts are presented in Figure \ref{fig:ctaIV}. An integration radius of 0.204 degrees was used in this scenario. The maximum detectable cutoff energy in the CTAO simulated spectrum is $\sim$ 13 TeV and 7 TeV for CTA III and CTA IV, respectively. The two simulations of the CTAO observations and the LHAASO observation are shown in Figure \ref{fig:6}. Despite a slight difference in flux amplitude, they can still be distinguished by the maximum cutoff energy detectable by CTAO.

In conclusion, Table \ref{tab:9} displays the spectral model parameters for the regions of the four pulsars. These findings demonstrate the effectiveness of CTAO in resolving these sources, which contributes significantly to the improved detection of PeVatrons in the Galaxy.

\begin{table}[htbp]
\centering
\begin{tabular}{llll}
\toprule
Source & $\rm{\Gamma}$ & $\rm{\phi_0\ (cm^{-2} s^{-1} TeV^{-1})}$ & $\rm{E_{cut}\ (TeV)}$    \\
\hline
\midrule
CTA I & 1.56 $\pm$ $<10^{-3}$ &        (\ 3.09 $\pm$ 0.01) $\times 10^{-14}$ & 75  $\pm$ 1 \\
CTA II & 2.47 $\pm$ $<10^{-3}$&        (\ 5.85 $\pm$ 0.01) $\times 10^{-14}$ & 299 $\pm$  12 \\
CTA III & 2.35 $\pm$ 0.09 &        (\ 9.21 $\pm$ 0.02) $\times 10^{-14}$& 13 $\pm$   $<1$  \\
CTA IV & 2.07 $\pm$  $< 10^{-3}$ &        (15.90 $\pm$ 0.04) $\times 10^{-14}$ &   7 $\pm$  $<1$ \\
\bottomrule
\end{tabular}
\caption{The parameters of the spectral model for the simulated CTAO observations.}
\label{tab:9}
\end{table}

\section{Gamma-ray as Galactic propagation probes CR} \label{sec:4}

This section describes the gamma emission originating from the PeVatron candidates LHAASO J1825-1326 and LHAASO J1839-0545. Considering a spin-down emission model, this emission is derived from a cosmic ray flux at the source. We compared the gamma flux obtained and the flux documented by the experiments, as discussed in the preceding section.

The diffusion/re-acceleration model is employed in our study. The simulation incorporates the influence of pointlike cosmic-ray sources, representing the practical implementation of pulsar sources. The calculations are performed inside a Cartesian coordinate system, where the Galactic plane is designated as the X-Y plane, and the Galactic Center (GC) is positioned at the origin. The spatial extent of the Galactic volume reaches a maximum of 20 kiloparsecs (kpc) in both the X and Y directions. Additionally, the Galactic volume has a halo height (h) of 8 kpc relative to the X-Y plane, where the Z-coordinate equals zero. To represent two-zone diffusion scenarios, we employ inhomogeneous diffusion features in the vicinity of CR sources, as detailed in \cite{2021JCAP...10..023D}.

Determining the spectrum energy distribution of the overall particles and gamma-ray emission involves utilizing source parameters. The model assumes accelerated particles are introduced into the interstellar medium (ISM). The spectral model for these injected particles follows a power-law distribution, expressed as $dq(p)/dp \propto p^{-\alpha}$, where $\alpha$ represents the spectral index and $q(p)$ the injection energy spectrum. Additionally, the total spectrum is scaled to ensure that the overall injected power is accurately represented by 
expression~\cite{2009PhRvD..80f3005M,2019ApJ...879...91J,Porter_2022}
\begin{equation}
L(t) = \eta L_{0}\Biggl(1 + \frac{t}{\tau_0}\Biggl)^{-2} ,
\label{power}
\end{equation}
where $\mathrm{{L_0}}$ represents the initial spin-down power of the pulsars and $L$ is the total particle emission from pulsars. The energy density in the vicinity of the pulsar's surface is primarily s influenced by the magnetic field, and the spin-down luminosity is predominantly attributed to magnetic dipole radiation. For most Pulsar Wind Nebulae, it is generally assumed that particles dominate the energy density. This suggests that at substantial distances from the neutron star, a significant fraction of the energy outflow has been transformed into particles, and the efficiency factor ($\eta$) during this phase, is $\eta = 1.0$. However, these particles do not immediately escape into the ISM but remain confined within the PWN due to its magnetic field, where they can experience significant energy loss \cite{2009PhRvD..80f3005M}. In this study, we adopt a conservative approach, considering $\eta$ to be approximately 0.1 to LHAASO J1839-0545, to account for the energy loss from particle cooling before their release into the ISM. Nonetheless, for LHAASO J1825-1326, given its high luminosity, we adjust $\eta$ to approximately 0.05. The pulsar time scale, $\tau_0$, is the ratio of the initial rotational energy to the initial spin-down luminosity \cite{2009PhRvD..80f3005M}. The results of the $\tau_0$ are summarized in Table~\ref{tab:tau}. 
\begin{table}[h!]
\centering
\begin{tabular}{ll}
\toprule
Pulsar &\ \  $\tau_0\  (\mathrm{yr})$  \\
\midrule
PSR J1826-1334 & 17.60 \\
PSR J1826-1256 & 20.62 \\
PSR J1837-0604 & 2.04 \\
PSR J1838-0537 & 3.37 \\
\bottomrule
\end{tabular}
\caption{Pulsar values for time scale $\tau_0$. \label{tab:tau}}
\end{table}
The initial spin-down power is calculated using the current spin-down power and the characteristic age of the pulsars. We assume that this emission $\mathrm{L}$ is caused by the rotational energy loss of the pulsar \cite{2009PhRvD..80f3005M}.
Within our framework, the diffusion coefficient is considered a scalar function that is uniform and isotropic across the Galaxy. It depends on the particle rigidity through a power law with an index $\delta$: $D_{xx} = \beta D_{0}(\rho/\rho_0)^{\delta}$, where $\rho_0 = 3$ GV, $D_0$ is a diffusion constant, and $\beta$ is the velocity in the unit of light speed. Models for the propagation of cosmic rays typically assume an isotropic diffusion coefficient to incorporate the random deflection of cosmic rays by the interstellar magnetic field \cite{Moskalenko_1998}. We considered for CR dynamics the following parameters: a diffusion constant of $D_{0}= 5.8\times 10^{28}$ $\mathrm{cm^{2}\ s^{-1}}$, a diffusion coefficient slope of $\delta = 0.33$, an Alfvén velocity of $v_{a} = 28.0$ $\mathrm{km\ s^{-1}}$, and spectral index of $\alpha = 2.8$. These values align with common diffusion coefficient values derived from fitting cosmic-ray data \cite{2007ARNPS..57..285S}. 

\begin{figure}[h!]
\centering
\subfloat[PSR J1826-1256]
{\includegraphics[angle=0,width=0.7\textwidth]{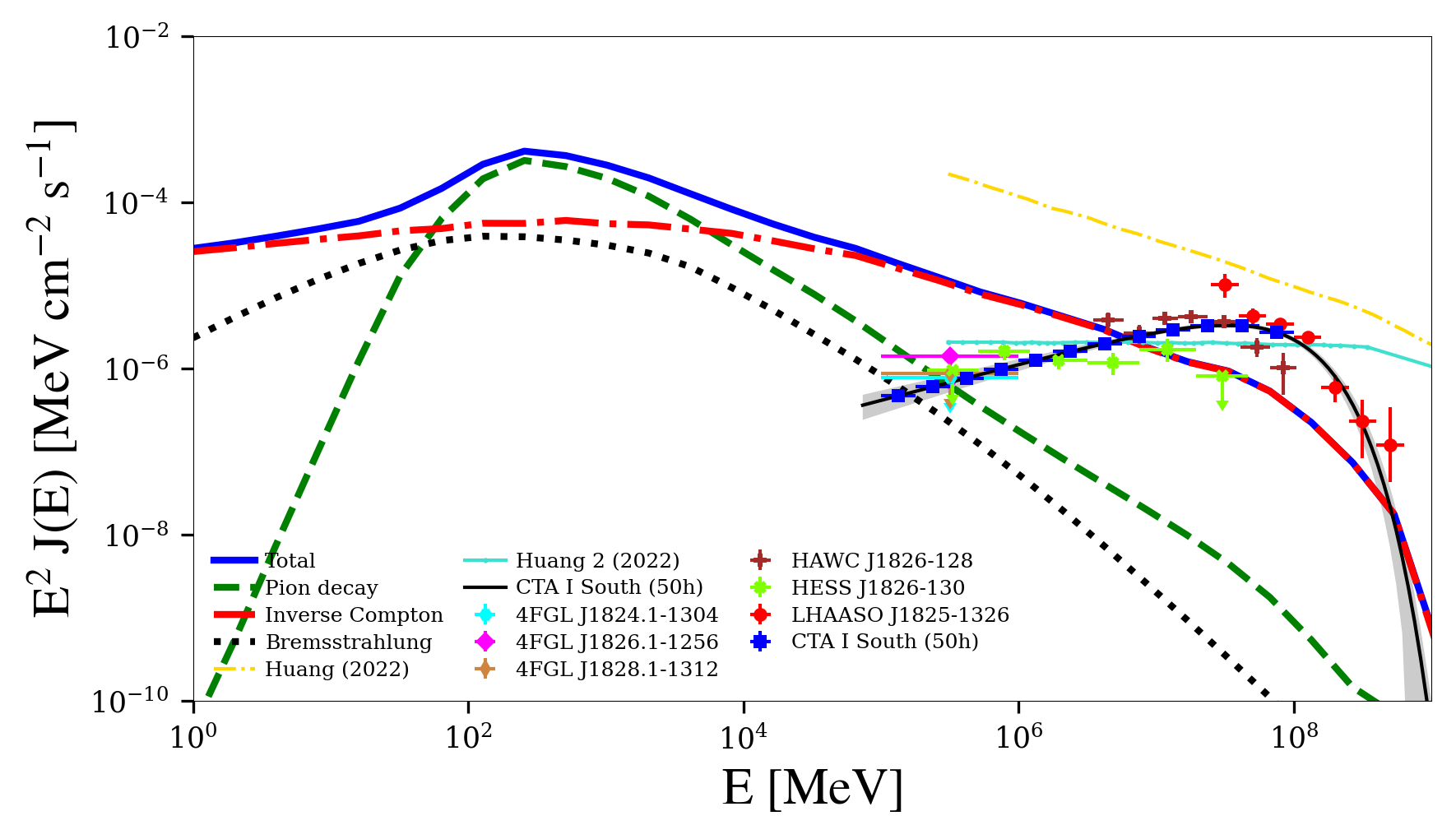}}\\
\subfloat[PSR J1826-1334]
{\includegraphics[angle=0,width=0.7\textwidth]{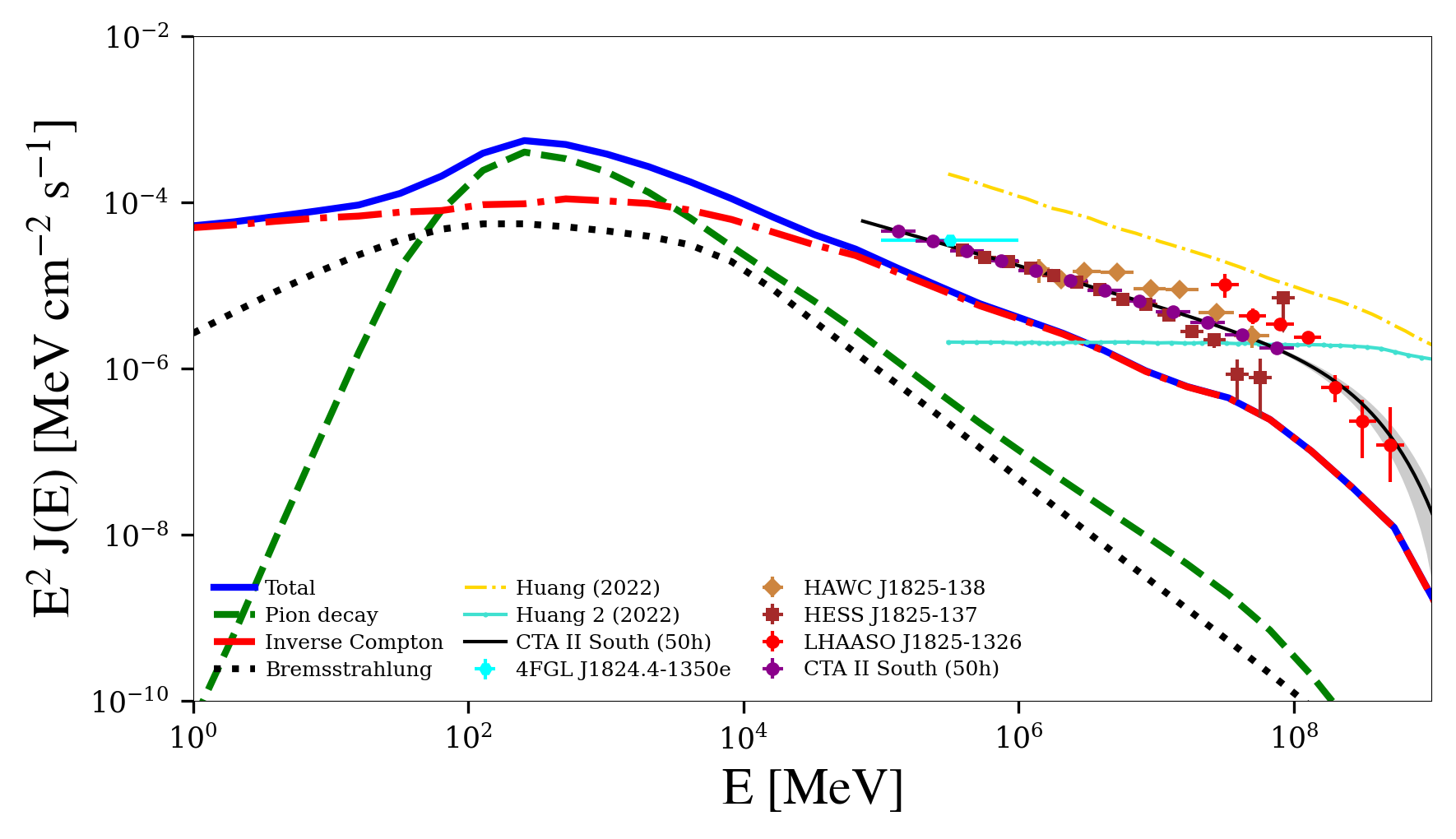}}
\caption{Spectral energy distribution of the total gamma-ray emission from PeVatron LHAASO J1825-1326. The total gamma-ray spectrum is the sum of hadronic and leptonic contributions.}
\label{fig:pevatron1}
\end{figure}

\begin{figure}[h!]
\centering
\subfloat[PSR J1837-0604]
{\includegraphics[angle=0,width=0.7\textwidth]{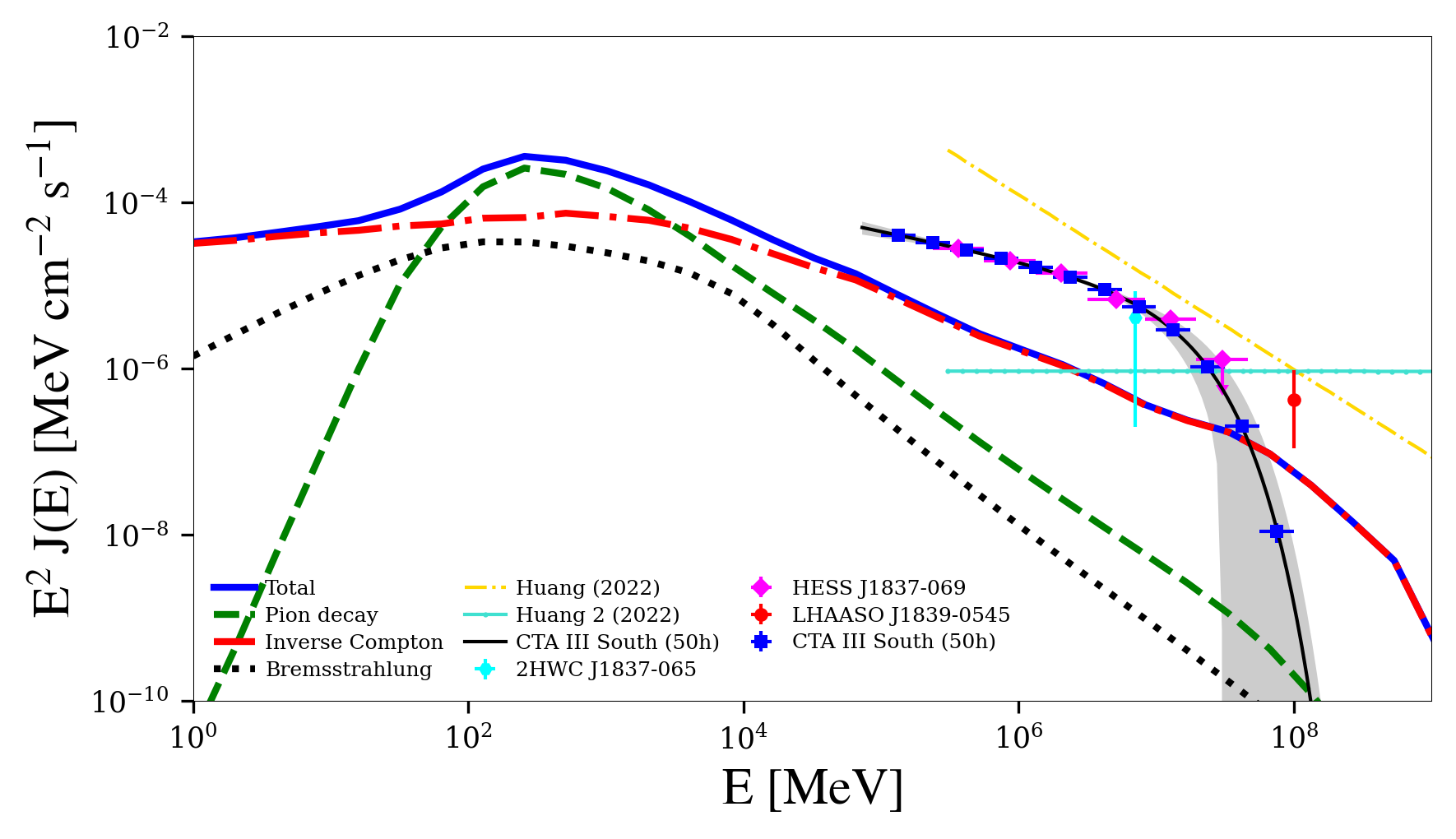}}\\
\subfloat[PSR J1838-0537]
{\includegraphics[angle=0,width=0.7\textwidth]{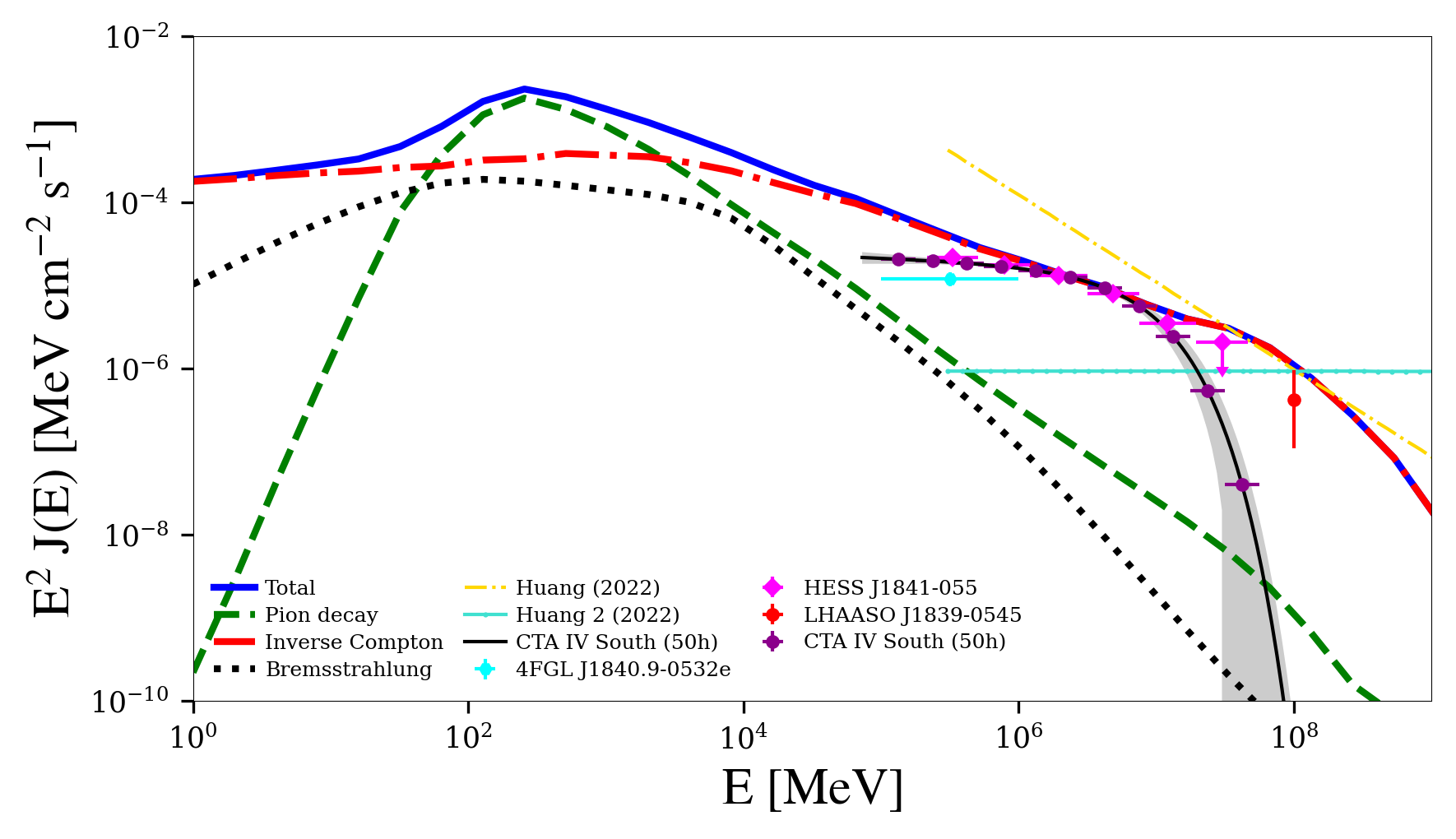}}
\caption{Spectral energy distribution of the total gamma-ray emission from PeVatron LHAASO J1839-0545. The total gamma-ray spectrum is the sum of hadronic and leptonic contributions.}
\label{fig:pevatron2}
\end{figure}

The outcomes of gamma-ray analysis are presented in Figures \ref{fig:pevatron1} and \ref{fig:pevatron2}. The figures also describes a spectral energy distribution of pulsars, analyzed using gamma-ray observations from CTAO simulations.   In the spin-down scenario, gamma-ray emission is mainly produced by Inverse Compton at high energies. The results show that the gamma-ray contribution from a hadronic origin to the highest energies is less than 1\%. Figure \ref{fig:pevatron1} depicts the findings for PSR J1826-1256 and PSR J1826-1334 within the PeVatron region LHAASO J1825-1326. The spectrum in Figure \ref{fig:pevatron1}-(a) illustrates a significant gamma-ray contribution from cosmic rays at high energies. On the other hand, the spectrum in Figure \ref{fig:pevatron1}-(b) shows a lower intensity contribution at high energies.
The same analysis was carried out in Figure \ref{fig:pevatron2} for PeVatron LHAASO J1839-0545. Simulations were conducted for PSR J1837-0604 and PSR J1838-0537 pulsars. As shown in Figure \ref{fig:pevatron2}-(a), there is a lower intensity contribution at high energies compared to Figure \ref{fig:pevatron2}-(b). These findings show that accelerated cosmic rays from the spin-down model can contribute to the gamma-ray flux at lower energies. The preliminary calculations reported here imply that cosmic rays contribute considerably to the measured total gamma radiation, despite the need for further improvement in the cosmic ray emission model at the source. 

\section{Discussion and concluding remarks}\label{sec:5}

This article comprehensively discusses two PeVatrons detected by the LHAASO. Each source is associated with regions containing two pulsars, which are conducive environments for particle acceleration. We assess the detection capabilities of CTAO for these PeVatrons through an ON/OFF analysis, and our analysis demonstrates the effectiveness of this approach. As a result of this analysis, we obtained spectral models for the four pulsars that are possible origins for the aforementioned PeVatrons. CTAO will be able to investigate the gamma-ray emission of these pulsars thoroughly.

Additionally, we investigated the potential of these pulsar environments as sources of cosmic-ray emissions and examined the resulting gamma radiation emissions. Our findings suggest that the cosmic-ray flux at the source can generate gamma emissions from pions and electrons up to energies of 100 TeV and 1000 TeV, respectively. However, for a more in-depth analysis of these results, we plan to explore the gamma emission at the source using multi-wavelength models in a future work, incorporating data that has already been measured. In the future, the CTAO \cite{CTACollaboration}, HAWC \cite{Abeysekara2017}, and LHAASO \cite{Zhen2010} experiments will enable the observation of photons with significantly higher energy levels. This would provide more stringent constraints on lepton and hadron acceleration at higher energy thresholds.

\acknowledgments

We are grateful to the anonymous Referee for the suggestions that helped us improve the manuscript. The research of RCA is supported by
Conselho Nacional de Desenvolvimento Científico e Tecnológico (CNPq) grant numbers
310448/2021-2 and 4000045/2023-0. She thanks AWS Cloud Credit for the Research. She also thanks for the support of L'Oreal Brazil, with the partnership of ABC and UNESCO in Brazil. We acknowledge the National Laboratory for Scientific Computing (LNCC/MCTI, Brazil) for providing HPC resources of the SDumont supercomputer, which have contributed to the research results reported within this paper: \href{http://sdumont.lncc.br}{http://sdumont.lncc.br}. R.C.A., R.J.C, D.B.G and A.J.T.S.M. acknowledge the financial support from the NAPI “Fenômenos Extremos do Universo” of Fundação de Apoio à Ciência, Tecnologia e Inovação do Paraná. R.C.A. also acknowledges Araucaria Foundation grant numbers
698/2022 and 721/2022; and FAPESP Project No. 2021/01089-1. This study was financed in part by the Coordenação de Aperfeiçoamento de Pessoal de Nível Superior – Brasil (CAPES) – Finance Code 001. This study utilized gamma-cat, an open data repository, and source catalog for gamma-ray astronomy, available at \href{https://github.com/gammapy/gamma-cat}{https://github.com/gammapy/gamma-cat}. The research employed the ATNF Pulsar Catalogue \cite{Manchester2005}, accessible at \href{https://www.atnf.csiro.au/research/pulsar/psrcat/}{https://www.atnf.csiro.au/research/pulsar/psrcat/}. The research also employed Gammapy, a Python package developed by the community for TeV gamma-ray astronomy \citep{Deil2017}, accessible at \href{https://www.gammapy.org}{https://www.gammapy.org}. In addition, we used the instrument response functions for the Cherenkov Telescope Array (CTA) provided by the CTA Consortium and Observatory. For detailed information about these instrument response functions, please refer to \href{https://www.ctao-observatory.org/science/cta-performance}{https://www.ctao-observatory.org/science/cta-performance} (version prod5 v0.1; \cite{CTAIRFS2021}).



\bibliographystyle{JHEP}
\bibliography{export}

\end{document}